\documentclass[aps,prl,reprint,showpacs]{revtex4-1} 
\usepackage{epsfig,amsmath,amssymb}
\usepackage{graphicx}
\usepackage[english]{babel}

\begin{document}

\title{ On the quantum-field description of many-particle \\ Bose
systems with spontaneously broken symmetry}
\author{Yu.\,M. Poluektov}
\email{yuripoluektov@kipt.kharkov.ua} %
\affiliation{%
National Science Center ``Kharkov Institute of Physics and
Technology'', 1, Akademicheskaya St., 61108 Kharkov, Ukraine }%

\begin{abstract}
A quantum-field approach to studying the Bose systems at finite
temperatures and in states with spontaneously broken symmetry, in
particular in a superfluid state, is proposed. A generalized model
of a self-consistent field (SCF) for spatially inhomogeneous
many-particle Bose systems is used as the initial approximation. A
perturbation theory has been developed, and a diagram technique for
temperature Green's functions (GFs) has been constructed. The
Dyson's equations joining the eigenenergy and vertex functions have
been deduced.
\end{abstract}
\pacs{ 05.30.Ch, 05.30.Jp, 05.70.-a } %
\maketitle

The application of quantum-field methods to the description of
interacting Bose particles meets the considerable difficulties. The
nature of these difficulties is associated with the fact that, at
sufficiently low temperatures, the Bose systems are in the state
with a with spontaneously broken phase symmetry. Therefore, one has
to utilize the quantum-field methods which have to be formulated
with regard for the symmetry breakdown. A success in the utilization
of the quantum-field perturbation theory essentially depends on the
correct choice of the zeroth approximation. As a rule, in the
standard approximation, the model of non-interacting particles is
used as an initial approximation, and the interaction Hamiltonian is
considered as a perturbation \cite{AGD}. Such a decomposition of the
Hamiltonian turns out to be inefficient under the utilization of
perturbation theory for the investigation of the systems with
spontaneously broken symmetry. Furthermore, if a model of the ideal
Bose gas with a condensate is chosen as the zero approximation, the
Wick's theorems, which are the basis of the perturbation theory and
diagram technique in a field theory, are inapplicable due to the
presence of the Bose condensate. However, S.T.\,Belyaev
\cite{Belyaev} managed to overcome the obstacles using the
N.N.\,Bogolyubov's idea \cite{Bogolyubov1} of a substitution of the
operators of particles with zero momentum by {\it c\,}-numbers. The
Belyaev's approach was further developed in Ref.\,\cite{HP}.
However, this approach is not sufficiently general. In particular,
it is not clear how the approach can be extended to spatially
inhomogeneous Bose systems, in which the Bose condensate contains
not only the particles with zero momentum, but also the particles
with nonzero one. What is more, the substitution of an operator by a
{\it c\,}-number, which is considered as a variational parameter, is
an approximation that essentially influences the theory structure.
Later on, in works \cite{NN,PS}, the attention was paid to the
paradoxicality of some results obtained within the frames of the
theory based on the model of ideal Bose gas. A modified variant of
the quantum-field theory \cite{N} developed to overcome the noted
difficulties contains a lot of assumptions and cannot be considered
as consistently microscopic.

The quantum-field description of the many-particle systems with
broken symmetry can be made more consistent by means of the
utilization of a SCF model as an initial approximation. For the case
of Fermi particles, a choice of such zero approximation for a
many-particle problem was proposed by Goldstone and Hubbard (see
references 2 and 9 in a book of collected articles \cite{PMQT}). A
description of the quantum-field methods constructed on the basis of
the SCF model is given in \cite{Kirzhnits}. What can be noted as a
remarkable property of the SCF equations is that they have the
solutions, whose symmetry is lower than that of the Hamiltonian of
the system. Thus, being formulated in a sufficiently general form,
the SCF equations can describe the states of many particles with
spontaneously broken symmetry. The SCF model for the spatially
inhomogeneous states of the Fermi systems with broken symmetry was
developed in \cite{Poluektov1}. The corresponding model for the Bose
systems was presented in \cite{Poluektov2}. The quantum-field
approach and diagram technique for the description of the Fermi
systems, which are in the states with broken symmetry at finite
temperatures, are formulated in \cite{Poluektov3,Poluektov4}.

This work is aimed at the development of a quantum-field approach
which, being based on the choice of the SCF model as an initial
approximation \cite{Poluektov2}, is able to describe the systems of
interacting Bose particles, which are in the states with the
spontaneously broken symmetry at finite temperatures. This approach
is founded only on the general principles of quantum mechanics and
statistical physics and requires no additional hypotheses. It can
also be used for the description of spatially inhomogeneous states
and is free from the difficulties of the approach based on the ideal
gas model.

{\bf 1.} The motion of a boson, whose spin is assumed to equal zero,
in the external field $U_0({\bf r})$ is described by the
Schr\"{o}dinger equation
\begin{equation} \label{EQ01}
\begin{array}{l}
\displaystyle{%
  \int\! dx' H_0(x,x')\,\varphi_j(x')=\varepsilon_j^{(0)}\varphi_j(x)\,, %
}%
\end{array}
\end{equation}
where the notation $x=\{{\bf r}\}$ is used. Index $j$ comprises the
full set of quantum numbers which characterize the stationary state
of an individual particle, $\varphi_j(x)$ is the wave function of
the particle, and $\varepsilon_j^{(0)}$ is its energy. The kernel in
Eq.\,(\ref{EQ01}) has the form
\begin{equation} \label{EQ02}
\begin{array}{c}
\displaystyle{%
  H_0(x,x')=-\frac{\hbar^2}{2m}\Delta\,\delta(x-x')+U_0({\bf r})\,\delta(x-x')\,,  %
}%
\end{array}
\end{equation}
where $m$ is the particle mass, and $\Delta$ is the Laplacian. Using
the secondary quantization apparatus, we introduce the operators of
creation, $a_j^+$, and annihilation, $a_j$, of a particle in the
state $j$ which obey the Bose commutation relations \cite{AGD}. We
also define the field operators
\begin{equation} \label{EQ03}
\begin{array}{l}
\displaystyle{%
  \Psi(x)=\sum_j \varphi_j(x)\,a_j, \quad  \Psi^+(x)=\sum_j \varphi_j^*(x)\,a_j^+\,. %
}%
\end{array}
\end{equation}
For the many-particle system under investigation, the Hamiltonian
expressed in terms of the field operators looks as
\begin{equation} \label{EQ04}
\begin{array}{ll}
\displaystyle{%
  H=\int\! dx\,dx'\,\Psi^+(x)H(x,x')\Psi(x')+  %
} \vspace{2mm}\\ %
\displaystyle{%
  \hspace{5mm}+\frac{1}{2}\int\! dx\,dx'\,\Psi^+(x)\Psi^+(x')U({\bf r},{\bf r}')\Psi(x')\Psi(x)\,,  %
}
\end{array}
\end{equation}
where $U({\bf r},{\bf r}')$ is the two-particle interaction
potential, and
\begin{equation} \nonumber
\begin{array}{ll}
\displaystyle{%
  H(x,x')=H_0(x,x')-\mu\,\delta(x-x')\,.  %
} \vspace{2mm}\\ %
\end{array}
\end{equation}
While studying the many-particle systems with broken symmetry, it is
convenient to assume that the system under consideration is in
contact with a thermostat and has the opportunity to exchange both
energy and particles with it, i.e. the total energy and the total
number of particles are supposed to be not fixed. The thermostat is
characterized by two parameters -- the temperature $T$ and the
chemical potential $\mu$. In the state of thermodynamic equilibrium,
the same parameters also characterize the system of particles. For
this reason, we use the grand canonical ensemble and will work with
the Hamiltonian that includes the term with the chemical potential
$-\mu N$, where $N$ is the operator of the number of particles.

{\bf 2.} At first, we formulate a general SCF model for the
Bose-systems with regard for the possibility of an arbitrary
breakdown of symmetry. It should be noted that a phenomenological
version of the SCF model, which is a generalization of the Fermi
liquid theory to the system of Bose particles, was developed in
works \cite{KP,AKPY,KP2}. To pass to the SCF model, we represent the
initial Hamiltonian Eq.\,(\ref{EQ04}) as the sum of two terms
\begin{equation} \label{EQ05}
\begin{array}{l}
\displaystyle{%
  H=H_0+H_C\,, %
}
\end{array}
\end{equation}
where the first term is the Hamiltonian of the SCF model, which
includes the terms with powers not higher than quadratic in the
field operators,
\begin{equation} \label{EQ06}
\begin{array}{ll}
\displaystyle{%
  H_0=\int\! dx\,dx'\biggl\{ \Psi^+(x)[H(x,x')+W(x,x')\Psi(x')]+
} \vspace{2mm}\\ %
\displaystyle{%
  \hspace{1mm} +\frac{1}{2}\Psi^+(x)\Delta(x,x')\Psi^+(x')\!+\!\frac{1}{2}\Psi(x')\Delta^*(x,x')\Psi(x)\!\biggr\}+ %
} \vspace{2mm}\\ %
\displaystyle{%
  \hspace{1mm} + \int\! dx\,[F(x)\Psi^+(x) + F^*(x)\Psi(x)] + E_0', %
}
\end{array}
\end{equation}
and the second one is the correlation Hamiltonian
\begin{equation} \label{EQ07}
\begin{array}{ll}
\displaystyle{%
  H_C=\frac{1}{2}\int\! dx\,dx'\biggl\{ \Psi^+(x)\Psi^+(x')U(x,x')\Psi(x')\Psi(x)- %
} \vspace{2mm}\\ %
\displaystyle{%
  \hspace{1mm} -2\Psi^+(x)W(x,x')\Psi(x')- %
} \vspace{2mm}\\ %
\displaystyle{%
  \hspace{1mm} -\Psi^+(x)\Delta(x,x')\Psi^+(x')-\Psi(x')\Delta^*(x,x')\Psi(x)   \!\biggr\} - %
} \vspace{2mm}\\ %
\displaystyle{%
  \hspace{1mm} - \int\! dx\,[F(x)\Psi^+(x) + F^*(x)\Psi(x)] - E_0', %
}
\end{array}
\end{equation}
which accounts for the particle correlations that are not included
in the SCF approximation. In contrast to the case of the Fermi
system \cite{Poluektov2,Poluektov3,Poluektov4},
Hamiltonian\,(\ref{EQ06}) of the SCF model contains also the terms
which are linear in the operators $\Psi$ and $\Psi^+$. Expressions
(\ref{EQ06}) and (\ref{EQ07}) contain the self-consistent potentials
$F(x), W(x,x')$, and $\Delta(x,x')$ which, being indefinite yet,
satisfy the conditions imposed by the Hamiltonian self-adjointness
\begin{equation} \label{EQ08}
\begin{array}{l}
\displaystyle{%
  W(x,x')=W^*(x',x), \,\,\, \Delta(x,x')=\Delta(x',x), %
}
\end{array}
\end{equation}
as well as the operator-free term $E_0'$, whose choice is essential
for the correct analysis of the thermodynamics within the model
under consideration. Thus, in the SCF model, Hamiltonian $H$
(\ref{EQ04}) is replaced by the simpler model Hamiltonian $H_0$
(\ref{EQ06}). The essential qualitative distinction between these
two Hamiltonians consists in that the initial Hamiltonian $H$ does
not depend on the system state, whereas the self-consistent one,
$H_0$, as will be shown below, depends on the system state and
thermodynamic variables through the self-consistent potentials
$F(x), W(x,x')$, and $\Delta(x,x')$. It is this property of the
self-consistent Hamiltonian that makes it possible to describe the
states with broken symmetry. To construct the perturbation theory
for the many-particle systems with broken symmetry, it is natural to
choose the self-consistent Hamiltonian $H_0$ as the basic one, and
the correlation Hamiltonian $H_C$ as a perturbation.

Hamiltonian (\ref{EQ06}) can be reduced to a diagonal form. To do
this, it is necessary to get rid of the terms which are linear in
Bose operators. We define the ``displaced'' Bose operators $\Phi(x)$
and $\Phi^+(x)$ as
\begin{equation} \label{EQ09}
\begin{array}{l}
\displaystyle{%
   \Psi(x)=\chi(x)+\Phi(x), \,\, \Psi^+(x)=\chi^*(x)+\Phi^+(x)\,. %
}
\end{array}
\end{equation}
The function $\chi(x)$ should be chosen in such a way that the
Hamiltonian $H_0$ wouldn't contain the terms linear in the field
operators. As a result, we obtain the condition
\begin{equation} \label{EQ10}
\begin{array}{l}
\displaystyle{%
   \int\! dx'\,[\Omega(x,x')\chi(x') + \Delta(x,x')\chi^*(x')] + F(x)=0\,, %
}
\end{array}
\end{equation}
where $\Omega(x,x')=H(x,x')+W(x,x')$. With regard for (\ref{EQ10}),
the Hamiltonian $H_0$ takes the form
\begin{equation} \label{EQ11}
\begin{array}{l}
\displaystyle{%
   H_0=\int\! dx\,dx'\biggl\{ \Phi^+(x)\,\Omega(x,x')\,\Phi(x')+
} \vspace{2mm}\\ %
\displaystyle{%
  \hspace{1mm} +\frac{1}{2}\Phi^+(x)\Delta(x,x')\Phi^+(x')\!+\!\frac{1}{2}\Phi(x')\Delta^*(x,x')\Phi(x)\!\biggr\}- %
} \vspace{2mm}\\ %
\displaystyle{%
  -\int\! dx\,dx'\biggl\{ \chi^*(x)\,\Omega(x,x')\,\chi(x') + %
} \vspace{2mm}\\ %
\displaystyle{%
  \hspace{1mm} +\frac{1}{2}\chi^*(x)\Delta(x,x')\chi^*(x')\!+\!\frac{1}{2}\chi(x')\Delta^*(x,x')\chi(x)\!\biggr\} + E_0'. %
}
\end{array}
\end{equation}
This Hamiltonian doesn't contain the terms which are linear in field
operators and can be reduced with the use of the Bogolyubov's
canonical transformations
\begin{equation} \label{EQ12}
\begin{array}{ll}
\displaystyle{%
  \Phi(x)= \sum_i \left[ u_i(x)\gamma_i + v_i^*(x)\gamma_i^+ \right]\,, %
} \vspace{1mm}\\ %
\displaystyle{%
  \Phi^+(x)= \sum_i \left[ v_i(x)\gamma_i + u_i^*(x)\gamma_i^+ \right]\,, %
}
\end{array}
\end{equation}
to the diagonal form
\begin{equation} \label{EQ13}
\begin{array}{ll}
\displaystyle{%
  H_0=E_0 + \sum_i \varepsilon_i\, \gamma_i^+\gamma_i\,, %
}
\end{array}
\end{equation}
where $E_0$ is the operator-free part of the Hamiltonian,
$\varepsilon_i$ -- the energy of elementary excitations,
quasiparticles, reckoned from the chemical potential, $i$ -- the
full set of quantum numbers characterizing the quasiparticle state.
The operators $\gamma_i^+$ and $\gamma_i$ describe the processes of
creation and annihilation of quasiparticles. The description in
terms of quasiparticles is widely used in condensed matter physics.
In the SCF model, the idea of quasiparticles, which possess the
infinite lifetime in this approximation, appears in a natural way as
a result of a reduction of Hamiltonian (\ref{EQ11}) to the diagonal
form (\ref{EQ13}). The relative simplicity of such a model consists
in the fact that it retains the the single-particle (to be precise,
single-quasiparticle) description of the system. The set of
coefficients $u(x)$ and $v(x)$ can be considered as the
two-component wave function of a quasiparticle. For the transition
from the self-consistent Hamiltonian (\ref{EQ11}) to the
diagonalized one (\ref{EQ13}) to be possible, the coefficients in
the canonical transformations (\ref{EQ12}) should satisfy the
Bogolyubov-de Gennes system of equations for the Bose systems
\cite{Poluektov2,Bogolyubov2,Gennes} which, in the most general
case, has the form
\begin{equation} \label{EQ14}
\begin{array}{ll}
\displaystyle{%
  \int\!dx'\left[ \Omega(x,x')\,u_i(x')+\Delta(x,x')\,v_i(x') \right]=\varepsilon_i\,u_i(x)\,,  %
} \vspace{2mm}\\ %
\displaystyle{%
  \int\!dx'\left[ \Omega^*(x,x')\,v_i(x')+\Delta^*(x,x')\,u_i(x') \right]=-\varepsilon_i\,v_i(x)\,.  %
}
\end{array}
\end{equation}
The requirement for transformations (\ref{EQ12}) to be canonical
leads to the conditions of normalization
\begin{equation} \label{EQ15}
\begin{array}{ll}
\displaystyle{%
  \int\!dx\left[ u_i(x)\,u^*_{i'}(x) - v_i(x)\,v^*_{i'}(x) \right]= \delta_{ii'} \,,  %
} \vspace{2mm}\\ %
\displaystyle{%
    \int\!dx\left[ u_i(x)\,v_{i'}(x) - v_i(x)\,u_{i'}(x) \right]= 0 \,,  %
}
\end{array}
\end{equation}
and completeness
\begin{equation} \label{EQ16}
\begin{array}{ll}
\displaystyle{%
  \sum_i \left[ u_i(x)\,u^*_{i}(x') - v_i^*(x)\,v_{i}(x') \right]= \delta(x-x') \,,  %
} \vspace{2mm}\\ %
\displaystyle{%
  \sum_i \left[ u_i(x)\,v_{i}^*(x') - v_i^*(x)\,u_{i}(x') \right]= 0 \,,  %
}
\end{array}
\end{equation}
of the solutions of the self-consistent equations (\ref{EQ14}).

The mean values of operators in the SCF model are expressed through
the normal $\tilde{\rho}$ and anomalous $\tilde{\tau}$
single-particle density matrices
\begin{equation} \label{EQ17}
\begin{array}{ll}
\displaystyle{%
  \tilde{\rho}(x,x')=\langle\Psi^+(x')\Psi(x)\rangle_0=  \rho(x,x')+\chi^*(x')\chi(x)\,, %
}\vspace{2mm}\\ %
\displaystyle{ \hspace{2mm}%
  \tilde{\tau}(x,x')=\langle\Psi(x')\Psi(x)\rangle_0=  \tau(x,x')+\chi(x')\chi(x)\,, %
}
\end{array}
\end{equation}
where the out-of-condensate density matrices have the form
\begin{equation} \label{EQ18}
\begin{array}{ll}
\displaystyle{%
  \rho(x,x')=\langle\Phi^+(x')\Phi(x)\rangle_0=   %
}\vspace{2mm}\\ %
\displaystyle{ \hspace{2mm}%
  =\sum_i [u_i(x) u_i^*(x') f_i + v_i^*(x) v_i(x') (1+f_i)]\,, %
}
\end{array}
\end{equation} \vspace{-5mm}
\begin{equation} \label{EQ19}
\begin{array}{ll}
\displaystyle{%
  \tau(x,x')=\langle\Phi(x')\Phi(x)\rangle_0=   %
}\vspace{2mm}\\ %
\displaystyle{ \hspace{2mm}%
  =\sum_i [u_i(x)v_i^*(x')\,f_i + v_i^*(x)u_i(x')(1+f_i)]\,. %
}
\end{array}
\end{equation}

The quasiparticle distribution function has the same form as in the
model of ideal Bose gas,
\begin{equation} \label{EQ20}
\begin{array}{ll}
\displaystyle{%
  f_i=\langle\gamma_i^+\gamma_i\rangle_0 = f(\varepsilon_i)=[\exp\beta\varepsilon_i-1]^{-1}\,,  %
}
\end{array}
\end{equation}
where $\beta=1/T$ is the reciprocal temperature. Since the
quasiparticle energy $\varepsilon_i$ is a functional of $f_i$,
formula (\ref{EQ20}) is a complicated nonlinear equation for the
distribution function. In Eqs.\,(\ref{EQ17})\,--\,(\ref{EQ20}), the
averaging is performed with the statistical operator
\begin{equation} \label{EQ21}
\begin{array}{ll}
\displaystyle{%
  \rho_0=\exp\beta (\Omega_0-H_0)\,,  %
}
\end{array}
\end{equation}
where the normalization constant
$\Omega_0=-T\ln[\textrm{Sp}\,e^{-\beta H_0}]$ is determined from the
condition $\textrm{Sp}\,\rho_0=1$ and represents the thermodynamic
potential of the system in the SCF model. The density matrices
(\ref{EQ18}) and (\ref{EQ19}), as well as $\tilde{\rho}(x,x')$ and
$\tilde{\tau}(x,x')$, satisfy the conditions
\begin{equation} \label{EQ22}
\begin{array}{ll}
\displaystyle{%
  \rho(x,x')=\rho^*(x',x), \,\,\, \tau(x,x')=\tau(x',x).   %
}
\end{array}
\end{equation}

Since, according to (\ref{EQ09}), the operators $\Phi(x)$ and
$\Phi^+(x)$ are linear in $\gamma, \gamma^+,$ and the Hamiltonian
$H_0$ (\ref{EQ13}) is quadratic, we have
\begin{equation} \label{EQ23}
\begin{array}{ll}
\displaystyle{%
  \langle\Phi(x)\rangle_0 = \langle\Phi^+(x)\rangle_0 = 0 %
}
\end{array}
\end{equation}
and, hence,
\begin{equation} \label{EQ24}
\begin{array}{ll}
\displaystyle{%
  \chi(x)=\langle\Psi(x)\rangle_0, \,\,\,  \chi^*(x)=\langle\Psi^+(x)\rangle_0\,. %
}
\end{array}
\end{equation}
It follows from (\ref{EQ24}) that, in the SCF model, $\chi(x)$ can
be considered as a wave function which determines the particle
number density in the single-particle Bose condensate. It is worth
to note that property (\ref{EQ23}) makes it handy to utilize the
operators $\Phi^+(x)$ and $\Phi(x)$ for the construction of the
perturbation theory. It is this point that makes the approach we
developed to be strongly different from the Belyaev's theory and its
modifications, where the overcondensate operators are determined in
such a way that their value averaged over an exact state of the
system turns into zero.

For the system of equations (\ref{EQ10}) and (\ref{EQ14}) to be
completely determined, the self-consistent potentials $F(x)$,
$W(x,x')$, and $\Delta(x,x')$ should be expressed in terms of the
functions $u(x)$, $v(x)$, and $\chi(x)$. This can be done provided
that the functional
\begin{equation} \label{EQ25}
\begin{array}{ll}
\displaystyle{%
  I=[\langle H-H_0 \rangle_0]^2  %
}
\end{array}
\end{equation}
achieves a minimum. The requirement for the minimality of functional
(\ref{EQ25}) implies that the potentials should be chosen to satisfy
the condition that the self-consistent Hamiltonian (\ref{EQ06})
approximates the the initial Hamiltonian (\ref{EQ04}) in the best
way. By varying functional (\ref{EQ25}) in the density matrices
(\ref{EQ17}), from the condition $\delta I=0$ we get the relation
between the self-consistent potentials and the complete
single-particle density matrices
\begin{equation} \label{EQ26}
\begin{array}{ll}
\displaystyle{%
  W(x,x')=U(x,x')\,\tilde{\rho}(x,x')\, +  %
}\vspace{2mm}\\ %
\displaystyle{ \hspace{2mm}%
  +\,\delta(x-x')\int\!\!dx''\,U(x,x'')\,\tilde{\rho}(x'',x'')\,,  %
}
\end{array}
\end{equation}
\vspace{-4mm}%
\begin{equation} \label{EQ27}
\begin{array}{ll}
\displaystyle{%
  \Delta(x,x')=U(x,x')\,\tilde{\tau}(x,x')\,.  %
}
\end{array}
\end{equation}
The variation of (\ref{EQ25}) in $\chi(x)$ under the condition
$\delta I=0$ leads to the expression
\begin{equation} \label{EQ28}
\begin{array}{ll}
\displaystyle{%
  F(x)=-2\chi(x)\int\!\!dx'\,U(x,x')\,|\chi(x')|^2\,.   %
}
\end{array}
\end{equation}
The substitution of Eqs.\,(\ref{EQ26})\,--\,(\ref{EQ28}) into
Eqs.\,(\ref{EQ10}) and (\ref{EQ14}) gives the closed system of
nonlinear integro-differential equations for the wave functions
$u(x)$, $v(x)$ and $\chi(x)$:
\begin{equation} \label{EQ29}
\begin{array}{ll}
\displaystyle{%
  \biggr[\! -\frac{\hbar^2}{2m}\Delta + U_0(x)-\mu +\int\!\!dx'\,U(x,x')\tilde{\rho}(x',x')\biggr]u_i(x) + %
}\vspace{2mm}\\ %
\displaystyle{ \hspace{0mm}%
  + \int\!\!dx' U(x,x')\!\big[ \tilde{\rho}(x,x')u_i(x')+\tilde{\tau}(x,x')v_i(x')\big]\!\!=\!\varepsilon_i u_i(x), %
}
\end{array}
\end{equation}
\begin{equation} \label{EQ30}
\begin{array}{ll}
\displaystyle{%
  \biggr[\! -\frac{\hbar^2}{2m}\Delta + U_0(x)-\mu +\int\!\!dx'\,U(x,x')\tilde{\rho}(x',x')\biggr]v_i(x) + %
}\vspace{2mm}\\ %
\displaystyle{ \hspace{0mm}%
  + \int\!\!dx' U(x,x')\!\big[ \tilde{\rho}^*(x,x')v_i(x')+\tilde{\tau}^*(x,x')u_i(x')\big]\!= %
}\vspace{2mm}\\ %
\displaystyle{ \hspace{68mm}%
  =\!-\varepsilon_i v_i(x), %
}
\end{array}
\end{equation}
\begin{equation} \label{EQ31}
\begin{array}{ll}
\displaystyle{%
  \biggr[\! -\frac{\hbar^2}{2m}\Delta + U_0(x)- \mu +  %
}\vspace{2mm}\\ %
\displaystyle{\hspace{3mm}%
  + \int\!\!dx'\,U(x,x')\big[\tilde{\rho}(x',x')-2|\chi(x')|^2\big]\biggr] \chi(x) + %
}\vspace{2mm}\\ %
\displaystyle{\hspace{3mm}%
  + \int\!\!dx'\,U(x,x')\big[\tilde{\rho}(x,x')\,\chi(x')+ \tilde{\tau}(x,x')\,\chi^*(x') \big]\! = 0.  %
}%
\end{array}
\end{equation}

Equations (\ref{EQ29})\,--\,(\ref{EQ31}) along with the conditions
(\ref{EQ15}) and (\ref{EQ16}) describe the many-particle Bose system
in the SCF approximation. The system of equations we obtained has
three types of solutions:
\begin{equation} \nonumber
\begin{array}{ll}
\displaystyle{%
  \textrm{I})\,\, \chi(x)=v_i(x)=0, \,\, u_i(x)\neq 0;  %
}\vspace{2mm}\\ %
\displaystyle{ \hspace{0mm}%
  \textrm{II})\,\, \chi(x)=0, \,\,\, v_i(x)\neq 0, \,\, u_i(x)\neq 0;  %
}\vspace{2mm}\\ %
\displaystyle{ \hspace{0mm}%
  \textrm{III})\,\, \chi(x)\neq 0, \,\,\, v_i(x)\neq 0, \,\, u_i(x)\neq 0.  %
}
\end{array}
\end{equation}
The first type of solutions describes the state in which the
symmetry with respect to the phase transformations
\begin{equation} \label{EQ32}
\begin{array}{ll}
\displaystyle{%
  \Psi(x) \rightarrow \Psi(x)\,e^{i\xi} %
}
\end{array}
\end{equation}
is not broken (here $\xi$ is an arbitrary phase). In this ``normal''
state, the system contains neither a single-particle nor pair
condensate and doesn't display superfluidity. The second type of
solutions describes the states which are characterized by the broken
symmetry with respect to transformation (\ref{EQ32}) due to the
creation of the pair condensate analogous to that which appears in
the superfluid Fermi systems \cite{BCS,BTS}. In this case, the Bose
system displays the superfluidity. The superfluidity of Bose
systems, which results from their pair correlations, was studied in
works \cite{KP2,GA,Kondratenko,NP}. The solutions of the third type
describe the superfluid states with broken phase symmetry, which
contain both the single-particle and pair Bose condensates. It is
worth to note that the solutions, for which
\begin{equation} \label{EQ33}
\begin{array}{ll}
\displaystyle{ \hspace{0mm}%
  \chi(x)\neq 0, \,\,\, v_i(x)= 0, \,\, u_i(x)\neq 0,  %
}
\end{array}
\end{equation}
do not exist. It is these solutions that correspond to the case of
ideal Bose gas below the Bose transition temperature, in which the
Bose condensate and the overcondensate particles coexist. Thus, the
system of non-interacting particles coexisting with the Bose
condensate and the system of interacting (even with an arbitrarily
small interaction) Bose particles with the broken phase symmetry are
two entirely distinct systems. It is the use of the model of ideal
gas with the condensate as a basic model that gives rise to the
difficulties on the construction of a consistent theory of the
many-particle Bose systems with broken symmetry \cite{NN,PS}. As is
seen, this is concerned with the fact that it is impossible to
describe the pair correlations, which always exist in the superfluid
systems of interacting particles, within the frames of the ideal gas
model. In the real superfluid Bose systems, the pair and higher
orders correlations, which break the phase symmetry, play the role
comparable with that of the single-particle Bose condensate. For
example, according to modern experimental estimations
\cite{BKKP,Kozlov}, only about $8\%$ of particles in the superfluid
$^4$He belong to the single-particle Bose-condensate, whereas the
remaining contribution to the superfluid density follows from the
pair and higher orders correlations.

In many cases, to calculate the equilibrium characteristics of the
system under investigation, it is enough to find the single-particle
density matrices; the calculation of the wave functions of
quasiparticles is not necessary. The system of equations for the
single-particle density matrices can be found from
Eqs.\,(\ref{EQ29}) and (\ref{EQ28}) and formulae
(\ref{EQ18}),\,(\ref{EQ19}). It can be written in the form
\begin{equation} \label{EQ34}
\begin{array}{ll}
\displaystyle{%
  -\frac{\hbar^2}{2m}(\Delta-\Delta')\,\tilde{\rho}(x,x')+[U_0(x)-U_0(x')]\,\tilde{\rho}(x,x')+  %
}\vspace{2mm}\\ %
\displaystyle{ \hspace{3mm}%
  +\int\!\!dx''\big[ U(x,x'')-U(x',x'')\big] \times   %
}\vspace{2mm}\\ %
\displaystyle{ \hspace{3mm}%
  \times\big[\tilde{\rho}(x,x'')\,\tilde{\rho}(x'',x')+ \tilde{\rho}(x,x')\,\tilde{\rho}(x'',x'') +  %
}\vspace{2mm}\\ %
\displaystyle{ \hspace{7mm}%
             \tilde{\tau}(x,x'')\,\tilde{\tau}^*\!(x'',x')-2\,\chi(x)\,\chi^*\!(x')\,|\chi(x'')|^2 \big]=0, %
}
\end{array}
\end{equation}
\vspace{-5mm}%
\begin{equation} \label{EQ35}
\begin{array}{ll}
\displaystyle{%
  -\frac{\hbar^2}{2m}(\Delta+\Delta')\,\tilde{\tau}(x,x')+  %
}\vspace{2mm}\\ %
\displaystyle{\hspace{3mm}%
  +\big[U_0(x)+U_0(x')+ U(x,x')-2\mu\big]\tilde{\tau}(x,x')\, +  %
}\vspace{2mm}\\ %
\displaystyle{ \hspace{3mm}%
  +\int\!\!dx'' \big[ U(x,x'')+U(x',x'')\big]\times  %
}\vspace{2mm}\\ %
\displaystyle{ \hspace{3mm}%
  \times \big[ \tilde{\rho}(x,x'')\,\tilde{\tau}(x'',x')+ \tilde{\rho}(x'',x'')\,\tilde{\tau}(x,x') +  %
}\vspace{2mm}\\ %
\displaystyle{ \hspace{4mm}%
  + \tilde{\rho}(x',x'')\,\tilde{\tau}(x'',x)- 2\,\chi(x)\,\chi(x')\,|\chi(x'')|^2 \big]=0.  %
}
\end{array}
\end{equation}
To these equations, we should add Eq.\,(\ref{EQ31}). It is enough to
know the overcondensate density matrices and the condensate wave
function in order to calculate the average of an arbitrary operator.

{\bf 3.} A distinctive feature of the SCF model, which should be
considered in the derivation of thermodynamic relations from
Hamiltonian\,(\ref{EQ06}), consists in that this Hamiltonian
contains the self-consistent potentials and the term which doesn't
include the operators depending on temperature and chemical
potential. To build a consistent SCF model and obtain the
thermodynamic relations, it is important to correctly choose the
operator-free term $E_0'$ in (\ref{EQ06}). Let us find it from the
condition $\partial I/\partial E_0'=0$ which is equivalent to the
condition of equality of the average values for the exact and
self-consistent Hamiltonians, $\langle H\rangle_0=\langle H_0\rangle_0$.
The result reads
\begin{equation} \label{EQ36}
\begin{array}{ll}
\displaystyle{%
  E_0'\!=\!-\frac{1}{2}\int\!dx\,dx'\,U(x,x')\langle\Psi^+(x)\Psi^+(x')\Psi(x')\Psi(x)\rangle_0+ %
}\vspace{2mm}\\ %
\displaystyle{ \hspace{8mm}%
  + 2\int\!dx\,dx'\,U(x,x')\,|\chi(x)|^2|\,\chi(x')|^2.  %
}
\end{array}
\end{equation}
Using the definitions of thermodynamic potential (\ref{EQ21}) and
entropy $S_0=-\textrm{Sp}(\rho_0\ln\rho_0)$, it is easy to make sure
that the thermodynamic relation $\Omega_0=E-TS_0-\mu N$
($E$ is the total energy of the system) is fulfilled, and the variation
of the thermodynamic potential is equal to the averaged variation of
$H_0$:
\begin{equation} \label{EQ37}
\begin{array}{ll}
\displaystyle{%
  \delta\Omega_0=\langle\delta H_0\rangle_0\,.  %
}
\end{array}
\end{equation}
Expressing the self-consist Hamiltonian through the functions
$\chi(x)$, $\rho(x,x')$ and $\tau(x,x')$
(or $\tilde{\rho}(x,x')$, $\tilde{\tau}(x,x')$)
and varying it with regard for (\ref{EQ37}), we obtain
\begin{equation} \label{EQ38}
\begin{array}{ll}
\displaystyle{%
 \frac{\delta\Omega_0}{\delta\chi^*(x)}\!=\!\bigg\langle\!\frac{\delta H_0}{\delta\chi^*(x)}\!\bigg\rangle_0  %
 \!=\!\frac{\delta\Omega_0}{\delta\rho(x,x')}\!=\!\bigg\langle\!\frac{\delta H_0}{\delta\rho(x,x')}\!\bigg\rangle_0\!=  %
}\vspace{2mm}\\ %
\displaystyle{ \hspace{8mm}%
 \!=\!\frac{\delta\Omega_0}{\delta\tau^*(x,x')}\!=\!\bigg\langle\!\frac{\delta H_0}{\delta\tau^*(x,x')}\!\bigg\rangle_0 = 0. %
}
\end{array}
\end{equation}
To be able to deal with the full density matrices, the substitutions
$\rho(x,x')\rightarrow \tilde{\rho}(x,x')$ and
$\tau(x,x')\rightarrow \tilde{\tau}(x,x')$ should be made in
(\ref{EQ38}). As is seen from (\ref{EQ38}, the relations between the
fields $F(x)$, $W(x,x')$ and $\Delta(x,x')$, on the one hand, and
the wave function of the condensate $\chi(x)$ and the
single-particle density matrices $\rho(x,x'), \tau(x,x')$, on the
other hand, which have been established with the use of the
variational principle, make the thermodynamic potential extremal
with respect to its variation in $\delta\chi, \delta\rho$ and
$\delta\tau$. As follows from (\ref{EQ38}), the ordinary
thermodynamic relation
\begin{equation} \label{EQ39}
\begin{array}{ll}
\displaystyle{%
 d\Omega_0=-S_0 dT - N d\mu %
}
\end{array}
\end{equation}
is fulfilled at a fixed volume. The total energy can be found
either by means of the direct averaging of the energy operator
or with the help of the thermodynamic relation in terms of the
thermodynamic potential:
\begin{equation} \label{EQ40}
\begin{array}{ll}
\displaystyle{%
  E=\Omega_0 -\mu\frac{\partial\Omega_0}{\partial\mu}-T\frac{\partial\Omega_0}{\partial T}\,.  %
}
\end{array}
\end{equation}
It follows from (\ref{EQ39}) and (\ref{EQ40}) that, although
the self-consistent Hamiltonian contains the potentials which
depend on thermodynamic variables, this doesn't lead to the
violation of the thermodynamic relations, as one could
suggest \cite{Kirzhnits}, and, therefore, the SCF approximation
in statistics is intrinsically non-contradictory.

The total number $N$ of the particles in the Bose system can
be written in the form
\begin{equation} \label{EQ41}
\begin{array}{ll}
\displaystyle{%
  N=\int\!dx\,\tilde{\rho}(x,x)=N_Q+N_B,  %
}
\end{array}
\end{equation}
where $N_Q=\int\!dx\,n_Q(x)$ and $N_B=\int\!dx\,|\chi(x)|^2$ are the
numbers of overcondensate particles and particles in the
single-particle condensate, respectively, and %
$n_Q(x)=\sum_i \big[ |u_i(x)|^2f_i + |v_i(x)|^2(1+f_i) \big]$. %
Taking (\ref{EQ17}) and (\ref{EQ18}) into account, we obtain
$N_Q=N_q+N_p$, where $N_q=\sum_i f_i$
is the number of quasiparticles, and
\begin{equation} \label{EQ42}
\begin{array}{ll}
\displaystyle{%
  N_p= \sum_i \int\!\!dx\,|v_i(x)|^2\,\textrm{cth}\frac{\beta\varepsilon_i}{2}  %
}
\end{array}
\end{equation}
can be considered as number of particles which take part in the formation
of the condensate of Cooper pairs in a Bose system. In the case of the state
with unbroken phase symmetry, the number of particles coincides with
the number of quasiparticles. On the contrary, in the case of the superfluid
state, where the phase symmetry is broken, the number of quasiparticles
is always smaller than that of particles, since the particles, which are
contained in the Bose condensate and in the condensate of Cooper pairs,
don't take part in the formation of quasiparticle excitations.
At zero temperature, the quasiparticle excitations completely vanish,
and all the particles belong to either the single-particle or pair condensate.

The total energy of the system of particles in the SCF approximation
can be represented as the sum of three contributions:
$E=E_1+E_2+E_3$, where $E_1$ is the energy of the particles
which are out of the single-particle condensate,
$E_2$ is the energy of the particles of the single-particle condensate,
and $E_3$ is the energy of the ``interaction'' of the condensate
and overcondensate particles. The first contribution can be written as
$E_1=T^{(1)}+U_E^{(1)}+U_D^{(1)}+U_{\textrm{ex}}^{(1)}+U_C^{(1)} $,
where
\begin{equation} \label{EQ43}
\begin{array}{ll}
\displaystyle{%
  T^{(1)}=-\frac{\hbar^2}{2m}\int\!\!dxdx'\,\delta(x-x')\,\Delta\rho(x,x') %
}
\end{array}
\end{equation}
is the kinetic energy of the particles which are out of the
single-particle condensate,
\begin{equation} \label{EQ44}
\begin{array}{ll}
\displaystyle{%
  U_E^{(1)}=\int\!\!dx\,U_0(x)\,n_Q(x) %
}
\end{array}
\end{equation}
is the energy of the out-of-condensate subsystem in an external field,
\begin{equation} \label{EQ45}
\begin{array}{ll}
\displaystyle{%
  U_D^{(1)}=\frac{1}{2}\int\!\!dxdx'\,U(x,x')\,n_Q(x)\,n_Q(x') %
}
\end{array}
\end{equation}
is the energy of the direct interaction between the out-of-condensate particles, %
\begin{equation} \label{EQ46}
\begin{array}{ll}
\displaystyle{%
  U_{\textrm{ex}}^{(1)}=\frac{1}{2}\int\!\!dxdx'\,U(x,x')\,|\rho(x,x')|^2 %
}
\end{array}
\end{equation}
is the energy of the exchange interaction between the out-of-condensate particles, %
and
\begin{equation} \label{EQ47}
\begin{array}{ll}
\displaystyle{%
  U_C^{(1)}=\frac{1}{2}\int\!\!dxdx'\,U(x,x')\,|\tau(x,x')|^2 %
}
\end{array}
\end{equation}
is the energy of the pair Bose condensate.

The energy of the single-particle condensate can be represented as a sum %
$E_2=T^{(2)}+U_E^{(2)}+U_D^{(2)}$, where
\begin{equation} \label{EQ48}
\begin{array}{ll}
\displaystyle{%
  T^{(2)}=-\frac{\hbar^2}{4m}\int\!\!dx \big[ \chi^*\!(x)\,\Delta\chi(x) + \chi(x)\,\Delta\chi^*\!(x) \big] %
}
\end{array}
\end{equation}
is the kinetic energy of the condensate,
\begin{equation} \label{EQ49}
\begin{array}{ll}
\displaystyle{%
  U_E^{(2)}=\int\!\!dx\,U_0(x)\,|\chi(x)|^2 %
}
\end{array}
\end{equation}
is the energy of the condensate in an external field, and
\begin{equation} \label{EQ50}
\begin{array}{ll}
\displaystyle{%
  U_D^{(2)}=\frac{1}{2}\int\!\!dxdx'\,U(x,x')\,|\chi(x)|^2\,|\chi(x')|^2 %
}
\end{array}
\end{equation}
is the energy of the interaction between the condensate particles.
The third contribution to the total energy is determined by the interaction
of the particles which are out of the condensate and those of the
single-particle condensate:
\begin{equation} \label{EQ51}
\begin{array}{ll}
\displaystyle{%
  E_3^{(2)}= \int\!\!dxdx'\,U(x,x')\times %
}\vspace{2mm}\\ %
\displaystyle{ \hspace{3mm}%
  \times\Big[ \rho(x,x')\,\chi^*\!(x)\,\chi(x') + n_Q(x)\,|\chi(x')|^2 +
}\vspace{2mm}\\ %
\displaystyle{ \hspace{3mm}%
  + \frac{1}{2}\tau(x,x')\,\chi^*\!(x)\,\chi^*\!(x') +
    \frac{1}{2}\tau^*\!(x,x')\,\chi(x)\,\chi(x') \Big].
}
\end{array}
\end{equation}
The thermodynamic potential of the Bose system can be written in the form
\begin{equation} \label{EQ52}
\begin{array}{ll}
\displaystyle{%
   \Omega_0= -\!\left(
   U_D^{(1)}+U_{\textrm{ex}}^{(1)}+U_C^{(1)}+U_D^{(2)}\right)-
}\vspace{2mm}\\ %
\displaystyle{ \hspace{3mm}%
  -\sum_i \varepsilon_i \int\!dx\, |v_i(x)|^2 - \int\!\!dxdx'\,U(x,x')\times   %
}\vspace{2mm}\\ %
\displaystyle{ \hspace{3mm}%
  \times\Big[ \rho(x,x')\,\chi(x)\,\chi^*\!(x') + n_Q(x)\,|\chi(x')|^2 +
}\vspace{2mm}\\ %
\displaystyle{ \hspace{3mm}%
  + \frac{1}{2}\tau(x,x')\,\chi^*\!(x)\,\chi^*\!(x') +
    \frac{1}{2}\tau^*\!(x,x')\,\chi(x)\,\chi(x') \Big]+
}\vspace{2mm}\\ %
\displaystyle{\hspace{3mm}%
   +T\sum_i\ln\left(1-e^{-\beta\varepsilon_i}\right).
}
\end{array}
\end{equation}

As in the case of ideal gas, the entropy is expressed in terms
of the quasiparticle distribution function as
\begin{equation} \label{EQ53}
\begin{array}{ll}
\displaystyle{%
  S_0= \sum_i \big[ (1+f_i)\ln(1+f_i) -f_i\ln f_i  \big].  %
}
\end{array}
\end{equation}
Since $f_i\rightarrow 0$ as $T\rightarrow 0$, it is obvious that
the entropy of the Bose system equals zero at the zero temperature.

{\bf 4.} Since the symmetry of the system state is lower than that
of its Hamiltonian, the conventional definition of an average cannot
be used while calculating theoretically the exact characteristics
observed in the systems with broken symmetry. At the same time, when
calculating the averages according to the ordinary rules of
statistical mechanics, the symmetry of the averages always coincides
with that of the Hamiltonian. Such contradiction does not arise in
the SCF model, because the system of self-consistent equations has
solutions with symmetry lower than that of the initial Hamiltonian.
To overcome the noted difficulties, Bogolyubov introduced the
conception of quasiaverages into statistical mechanics
\cite{Bogolyubov3}. According to this conception, for the states
with broken symmetry, the averages should be calculated not using
Hamiltonian (\ref{EQ04}) but a Hamiltonian which differs from
(\ref{EQ04}) by the terms that break its symmetry in an appropriate
way. In the framework of such an approach, however, some uncertainty
in the fields that violate symmetry remains. Since a choice of these
fields does not depend on interparticle interactions, it can turn
out that the interactions do not allow the existence of the states
possessing the symmetry which is imposed by the introduced field. In
work \cite{Poluektov5} it was proposed to determine the
quasiaverages using the self-consistent Hamiltonian as an addition
that violates the symmetry. In this case, the system can possess
only such symmetry which is allowed by interparticle interactions.

Although the symmetry of the Hamiltonians $H_0$ and $H_C$,
which depend on the system state, can be lower than that of the
initial Hamiltonian, it is natural that the symmetry of $H$
doesn't depend on the way how it is split and, thus, remains unchanged.
Therefor, in order to describe the systems with broken symmetry,
we introduce a more general Hamiltonian
\begin{equation} \label{EQ54}
\begin{array}{ll}
\displaystyle{%
  H_g = H_0 + g H_C  %
}
\end{array}
\end{equation}
which depends on a real parameter $g$. It is obvious that this
Hamiltonian coincides at $g=1$ with the initial one (\ref{EQ04}),
whereas it turns into the self-consistent Hamiltonian (\ref{EQ06})
at $g=0$. The variation of this parameter from zero to unity means
the inclusion of the correlation interaction. If $g$ is very close
to unity, Hamiltonian (\ref{EQ54}) almost coincides with the initial
one (\ref{EQ04}). However, the most important difference consists in
the fact that its symmetry coincides with that of the
self-consistent Hamiltonian and can be lower than the symmetry of
the initial Hamiltonian. Let us define the statistical operator
\begin{equation} \label{EQ55}
\begin{array}{ll}
\displaystyle{%
  \rho_g = e^{\beta(\Omega_g-H_g)}, %
}
\end{array}
\end{equation}
where $\Omega_g=-T\ln\big(\textrm{Sp}\,e^{-\beta H_g}\big)$.
We write the quasiaverage value of an arbitrary operator $A$
in the form
\begin{equation} \label{EQ56}
\begin{array}{ll}
\displaystyle{%
  \langle A\rangle = \lim_{g\rightarrow 1}\lim_{V\rightarrow\infty}\textrm{Sp}\,\rho_g A\,. %
}
\end{array}
\end{equation}
At certain values of the thermodynamic variables $\mu$ and $T$,
quasiaverages (\ref{EQ56}) can differ from the averages defined in
an ordinary way and, thus, can describe the states with broken
symmetry. From the mathematical point of view, a possible divergence
between averages and quasiaverages consists, as known
\cite{Bogolyubov3,AP}, in the dependence of the result on the order
of the transitions to the limit in Eq.\,(\ref{EQ56}). The passage to
the limit of the ``coupling constant'' $g$ should be carried out
after the thermodynamic passage to the limits $V\rightarrow\infty$
and $N\rightarrow\infty$, provided $N/V=\textrm{const}$. If the
symmetry isn't broken, quasiaverages (\ref{EQ56}) are identical to
the relevant conventional averages.

{\bf 5.} The correlation Hamiltonian (\ref{EQ07}) chosen as a
perturbation has a rather complicated structure. However, it can be
written in a more compact form with the use of the notion of the
normal product of operators. The relations of perturbation theory
will take a simpler form in this case. This notion also plays the
essential role in quantum field theory. In the temperature-involved
technique \cite{AGD}, the notion of normal product isn't used,
therefore, the analogy with quantum field theory is incomplete.

For further consideration, it is convenient to introduce the
notation of operators using the ``isotopic'' index $\alpha$ which
takes two values, $1$ and $2$:
\begin{equation} \label{EQ57}
\begin{array}{ll}
\displaystyle{%
  a_{\alpha j}\!=\!\left\{\!
               \begin{array}{l}
                 a_j, \\
                 a_j^+,
               \end{array} \right. \,\,
  \gamma_{\alpha i}\!=\!\left\{\!
               \begin{array}{l}
                 \gamma_i, \\
                 \gamma_i^+,
               \end{array} \right. \,\,
  \Psi_\alpha(x)\!=\!\left\{\!
               \begin{array}{l}
                 \Psi(x), \\
                 \Psi^+(x),
               \end{array} \right.
}\vspace{2mm}\\ %
\displaystyle{\hspace{0mm}%
    \Phi_\alpha(x)\!=\!\left\{\!
    \begin{array}{l}
       \Phi(x), \\
       \Phi^+(x),
    \end{array} \right. \,\,
    \chi_\alpha(x)\!=\!\left\{\!
    \begin{array}{l}
        \chi(x), \\
        \chi^*\!(x),
    \end{array} \right. \,\,\,\,
\begin{array}{l}
  \alpha =1, \\
  \alpha =2.
\end{array}
}
\end{array}
\end{equation}
The complete $\Psi$ and overcondensate $\Phi$ field operators are
connected by relation (\ref{EQ09})
\begin{equation} \label{EQ58}
\begin{array}{ll}
\displaystyle{%
  \Psi_\alpha(x)=\chi_\alpha(x)+\Phi_\alpha(x). %
}
\end{array}
\end{equation}
We introduce also a notation
\begin{equation} \label{EQ59}
\begin{array}{ll}
\displaystyle{%
  \bar{\alpha}=\left\{
               \begin{array}{l}
                 1, \textrm{ when }\hspace{1mm} \alpha=2, \\
                 2, \textrm{ when }\hspace{1mm} \alpha=1.
               \end{array} \right.
}%
\end{array}
\end{equation}
We now give a general definition, valid for both the Fermi and Bose
statistics, for the normal product of operators \cite{Poluektov3,Poluektov4}.
We introduce the notion of the operator pairing which implies the averaging
over the self-consistent state:
\begin{equation} \label{EQ60}
\begin{array}{ll}
\displaystyle{%
  \eta_1^a\eta_2^a = \langle\eta_1\eta_2\rangle_0\,.  %
}
\end{array}
\end{equation}
Here, $\eta_i$ is any of the operators $a_{\alpha j}, \Phi_\alpha$
or $\gamma_{\alpha i}$. The product of an arbitrary number of operators
containing the pairings is defined as
\begin{equation} \label{EQ61}
\begin{array}{ll}
\displaystyle{%
  \eta_1^a\eta_2\eta_3^a\eta_4\ldots\eta_k^b\ldots\eta_m^b\ldots\eta_{j-1}\eta_j =  %
}\vspace{2mm}\\ %
\displaystyle{ \hspace{2mm}%
  =a\langle\eta_1\eta_3\rangle_0 \langle\eta_k\eta_m\rangle_0\times %
}\vspace{2mm}\\ %
\displaystyle{ \hspace{2mm}%
  \times\, \eta_2\eta_4\ldots \eta_{k-1}\eta_{k+1}\ldots \eta_{m-1}\eta_{m+1}\ldots\eta_{j-1}\eta_j\,, %
}%
\end{array}
\end{equation}
where $a$ is the multiplier which equals unity for the Bose
operators and $(-1)^p$ for the Fermi ones. Here, $p$ is the number of
permutations necessary to arrange the operators, which are paired,
side by side in the initial order. With regard for the given definition
of pairings, the normal product of any number of operators is
determined as
\begin{equation} \label{EQ62}
\begin{array}{ll}
\displaystyle{%
  N(\eta_1\eta_2\ldots\eta_j)=\eta_1\eta_2\ldots\eta_j- %
}\vspace{2mm}\\ %
\displaystyle{ \hspace{3mm}%
  - \eta_1^a\eta_2^a\eta_3\ldots\eta_j - \eta_1^a\eta_2\eta_3^a\ldots\eta_j - %
}\vspace{2mm}\\ %
\displaystyle{ \hspace{3mm}%
  - (\textrm{all other products with single pairing}) +  %
}\vspace{2mm}\\ %
\displaystyle{ \hspace{3mm}%
  + \eta_1^a\eta_2^a\eta_3^b\eta_4^b\ldots\eta_j + \eta_1^a\eta_2^b\eta_3^a\eta_4^b\ldots\eta_j +  %
}\vspace{2mm}\\ %
\displaystyle{ \hspace{3mm}%
  + (\textrm{all other products with two pairings}) - \ldots \,\,\,.  %
}
\end{array}
\end{equation}
Thus, the temperature normal product of operators is determined as
the sum of the products of operators which contain all possible pairings
(including a term without pairings). If the number of the pairings
in a product is even, the sign plus should be chosen in front of the term.
If the number of the pairings is odd, we should take the sign minus.
Let us consider the $N$-product of an arbitrary quantity of the operators
taken in either the Schr\"{o}dinger or interaction representation.
Its average, which is calculated over a self-consistent state,
equals zero, i.e.
\begin{equation} \label{EQ63}
\begin{array}{ll}
\displaystyle{%
  \big\langle N(\Psi_1\ldots\Psi_j\big\rangle_0 = 0\,,  %
}
\end{array}
\end{equation}
except for the case of the average of the $N$-product of $c$-numbers
which is $N(c)=c$ by definition.

The sufficiently complicated correlation Hamiltonian (\ref{EQ07})
can be written in terms of overcondensate operators and density matrices as
\begin{equation} \label{EQ64}
\begin{array}{ll}
\displaystyle{%
  H_C=\frac{1}{2}\int\!\!dxdx'\,U(x,x')\Big[ \Phi^+(x)\Phi^+(x')\Phi(x')\Phi(x)-  %
}\vspace{2mm}\\ %
\displaystyle{ \hspace{2mm}%
  - 2\rho(x,x')\Phi^+(x)\Phi(x')-2\rho(x',x')\Phi^+(x)\Phi(x)- %
}\vspace{2mm}\\ %
\displaystyle{ \hspace{2mm}%
  - \tau(x,x')\Phi^+(x)\Phi^+(x')-\tau^*\!(x,x')\Phi(x')\Phi(x)+  %
}\vspace{2mm}\\ %
\displaystyle{ \hspace{2mm}%
  + \rho(x,x')\rho(x',x)+\rho(x,x)\rho(x',x')+\tau(x',x)\tau^*\!(x',x)+  %
}\vspace{2mm}\\ %
\displaystyle{ \hspace{2mm}%
  + 2\chi^*\!(x)\Phi^+(x')\Phi(x')\Phi(x)+2\chi(x)\Phi^+(x)\Phi^+(x')\Phi(x')-  
}\vspace{2mm}\\ %
\displaystyle{ \hspace{2mm}%
  -2\rho(x,x')\chi(x')\Phi^+(x)-2\rho^*\!(x,x')\chi^*\!(x')\Phi(x)-
}\vspace{2mm}\\ %
\displaystyle{ \hspace{2mm}%
  -2\rho(x',x')\chi(x)\Phi^+(x)-2\rho(x',x')\chi^*\!(x)\Phi(x)-                
}\vspace{2mm}\\ %
\displaystyle{ \hspace{2mm}%
  -2\tau(x,x')\chi^*\!(x')\Phi^+(x)-2\tau^*\!(x,x')\chi(x')\Phi(x) \Big]
}
\end{array}
\end{equation}
As an important property of the SCF model, we note that it allows us
to represent the above Hamiltonian as the normal product of the
field operators. The sufficiently bulky correlation Hamiltonian
(\ref{EQ64}) consists of two terms:
\begin{equation} \label{EQ65}
\begin{array}{ll}
\displaystyle{%
  H_C=H_C^{(3)}+H_C^{(4)},  %
}
\end{array}
\end{equation}
where
\begin{equation} \label{EQ66}
\begin{array}{ll}
\displaystyle{%
  H_C^{(3)}= \int\!dx\,dx'\,U(x,x')\times %
}\vspace{1mm}\\ %
\displaystyle{ \hspace{0mm}%
  \times\!\Big[ \chi^*\!(x) N[\Phi^+\!(x')\Phi(x')\Phi(x)]\chi(x)N[\Phi^+\!(x)\Phi^+\!(x')\Phi(x')]\Big],  %
}\vspace{3mm}\\ %
\displaystyle{ \hspace{0mm}%
   H_C^{(4)}= \int\!dx\,dx'\,U(x,x')\,N[\Phi^+\!(x)\Phi^+\!(x')\Phi(x')\Phi(x)].  %
}
\end{array}
\end{equation}
The Hamiltonian $H_C^{(3)}$ contains the normal products of three
operators multiplied by the wave function of the Bose condensate,
whereas the Hamiltonian $H_C^{(4)}$ contains the normal product of
four operators and doesn't contain the wave function of the Bose
condensate. We pay attention to the fact that, due to the intrinsic
property of a normal product, the averages of the correlation
Hamiltonians (\ref{EQ66}) over the self-consistent state are equal
to zero:
\begin{equation} \label{EQ67}
\begin{array}{ll}
\displaystyle{%
  \left\langle H_C^{(3)}\right\rangle_0 = \left\langle H_C^{(4)}\right\rangle_0 =0. %
}
\end{array}
\end{equation}
On the construction of the perturbation theory, the correlation Hamiltonian
can be expressed through operators in the interaction representation as
\begin{equation} \label{EQ68}
\begin{array}{ll}
\displaystyle{%
  \Phi_\alpha(x,\tau)=e^{\tau H_0}\Phi_\alpha(x)\,e^{-\tau H_0},\,\,  %
  \gamma_{\alpha i}(\tau)=e^{\tau H_0}\gamma_{\alpha i}\,e^{-\tau
  H_0},
}%
\end{array}
\end{equation}
where $0\le\tau\le\beta$ is the Matsubara ``time'' parameter \cite{AGD}.
Since the Hamiltonian itself is integrated with respect to the time
variable, we can write
\begin{equation} \label{EQ69}
\begin{array}{ll}
\displaystyle{%
  \int_0^\beta \!d\tau H_C^{(3)}\!(\tau)\!=\!\frac{1}{2}\int\!d1\,d2\,\tilde{U}(1,2)\,\chi(1)\,N[\Phi(2)\Phi(\bar{2})\Phi(\bar{1})],  %
}\vspace{2mm}\\ %
\displaystyle{ \hspace{0mm}%
  \int_0^\beta \!d\tau H_C^{(4)}\!(\tau)\!=\!\frac{1}{8}\int\!d1\,d2\,\tilde{U}(1,2)\,N[\Phi(1)\Phi(2)\Phi(\bar{2})\Phi(\bar{1})], %
}
\end{array}
\end{equation}
where $1\!=\!(x_1,\tau_1,\alpha_1),
\bar{1}\!=\!(x_1,\tau_1,\bar{\alpha}_1)$, and so on. The integration
over a numerical variable means the integration over all continuous
variables and the summation over all discrete ones. In (\ref{EQ69}),
we introduced a symmetrized potential
\begin{equation} \label{EQ70}
\begin{array}{ll}
\displaystyle{%
  \tilde{U}(1,2)=\tilde{U}(x_1\tau_1\alpha_1,x_2\tau_2\alpha_2)=
}\vspace{2mm}\\ %
\displaystyle{ \hspace{0mm}%
  =U\!(x_1,x_2)\,\delta(\tau_1-\tau_2)(\delta_{\alpha_1\alpha_2}+\delta_{\alpha_1\bar{\alpha}_2}),
}
\end{array}
\end{equation}
whose symmetry properties are given by the relations
\begin{equation} \label{EQ71}
\begin{array}{ll}
\displaystyle{%
  \tilde{U}(1,2)=\tilde{U}(2,1)=\tilde{U}(\bar{1},2)=\tilde{U}(1,\bar{2})=\tilde{U}(\bar{1},\bar{2}).
}
\end{array}
\end{equation}
The correlation Hamiltonians expressed in terms of the quasiparticles operators
for both the Schr\"{o}dinger and interaction representations have the form
\begin{equation} \label{EQ72}
\begin{array}{ll}
\displaystyle{%
  H_C^{(3)}= \frac{1}{3!}\sum_{123}\left( \tilde{U}_{123}+\tilde{U}_{\bar{1}\bar{2}\bar{3}}^* \right)N(\gamma_1\gamma_2\gamma_3), %
}\vspace{2mm}\\ %
\displaystyle{ \hspace{0mm}%
  H_C^{(4)}= \frac{1}{4!}\sum_{1234} \tilde{U}_{1234}\,N(\gamma_1\gamma_2\gamma_3\gamma_4). %
}
\end{array}
\end{equation}
Each number in (\ref{EQ72}) denotes a collection of indices:
$1\!=\!(i_1,\alpha_1)$, $\bar{1}\!=\!(i_1,\bar{\alpha}_1)$, and so on.
The symmetrized matrix elements in (\ref{EQ72}) are expressed in terms
of the matrix elements
\begin{equation} \label{EQ73}
\begin{array}{ll}
\displaystyle{%
  U_{123}=U_{\,\,\,i_1i_2i_3}^{\alpha_1\alpha_2\alpha_3} = %
}\vspace{1mm}\\ %
\displaystyle{ \hspace{3mm}%
  = \int\!dx\,dx'\,U(x,x')\,\chi(x)\,u_{i_1}^{1\alpha_1}\!(x')\,u_{i_2}^{2\alpha_2}\!(x')\,u_{i_3}^{2\alpha_3}\!(x),  %
}\vspace{2mm}\\ %
\displaystyle{ %
  U_{1234}=U_{\,\,\,\,i_1i_2i_3i_4}^{\alpha_1\alpha_2\alpha_3\alpha_4} = %
}\vspace{1mm}\\ %
\displaystyle{ \hspace{3mm}%
  = \int\!dx\,dx'\,U(x,x')\,u_{i_1}^{2\alpha_1}\!(x)\,u_{i_2}^{2\alpha_2}\!(x')\,u_{i_3}^{1\alpha_3}\!(x')\,u_{i_4}^{1\alpha_4}\!(x)  %
}
\end{array}
\end{equation}
by the formulae
\begin{equation} \label{EQ74}
\begin{array}{ll}
\displaystyle{%
  \tilde{U}_{123}=U_{123}+U_{132}+U_{213}+U_{231}+U_{312}+U_{321},
}\vspace{3mm}\\ %
\displaystyle{ %
   \tilde{U}_{1234}\!=\!U_{1234}+U_{1243}+U_{1324}+U_{1342}+U_{1423}+U_{1432}+
}\vspace{2mm}\\ %
\displaystyle{\hspace{9mm}
   +U_{2314}+U_{2341}+U_{2413}+U_{2431}+U_{3412}+U_{3421}.
}
\end{array}
\end{equation}
The functions that determine the matrix elements in (\ref{EQ73}) are
expressed in terms of the coefficients of the Bogolyubov
transformation (\ref{EQ12}):
$u_i^{11}(x)\!=\!u_i^{22\,*}(x)\!=\!u_i(x)$, %
$v_i^{21}(x)\!=\!v_i^{12\,*}(x)\!=\!v_i(x)$. %
For the matrix elements depending on four indices, the symmetry
properties
\begin{equation} \label{EQ75}
\begin{array}{ll}
\displaystyle{%
  U_{1234}\!=\!U_{2143}\!=\!U_{\bar{4}\bar{3}\bar{2}\bar{1}}^*\!=\!U_{\bar{3}\bar{4}\bar{1}\bar{2}}^*,  %
  \textrm{\,\,\,as well as\,\,\,} \tilde{U}_{\bar{1}\bar{2}\bar{3}\bar{4}}^*\!=\!\tilde{U}_{1234}
}
\end{array}
\end{equation}
are fulfilled. As a result, only 7 of the 16 matrix elements of
$U_{1234}$, which differ from one another only by different
collections of isotopic indices $\alpha_i$, are independent ones and
enter into the correlation Hamiltonian in the form of three
combinations. There are 8 independent matrix elements of $U_{123}$
which differ from one another only by different collections of
isotopic indices $\alpha_i$ that enter into the correlation
Hamiltonian in the form of two combinations.

{\bf 6.} We define an arbitrary $L$-point temperature GF as %
\begin{equation} \label{EQ76}
\begin{array}{ll}
\displaystyle{%
  G(1,2,\ldots L)= i^L \Big\langle T_\tau \hat{A}(1)\hat{A}(2)\ldots \hat{A}(L)\Big\rangle\,,  %
}%
\end{array}
\end{equation}
where the averaging means the operation of quasiaveraging
(\ref{EQ56}), and each number stands for a whole set of variables.
The operators averaged in (\ref{EQ76}) are taken in the
Heisenberg-Matsubara representation as
\begin{equation} \label{EQ77}
\begin{array}{ll}
\displaystyle{%
  \hat{A}_\alpha(\tau)= e^{\tau H_g}A_\alpha\,e^{-\tau H_g}\,,  %
}%
\end{array}
\end{equation}
where $A_\alpha$ is an operator in the Schr\"{o}dinger
representation, $0\le\tau\le\beta$, and $T_\tau$ is the operator of
chronological ordering \cite{AGD}. Equation (\ref{EQ76}) determines
the $L$-point field GF if $\hat{A}(1)\!=\!\hat{\Psi}(1)$ and the
$L$-point quasiparticle GF if $\hat{A}(1)\!=\!\hat{\gamma}(1)$. For
the Fermi systems, the GFs are considered only with even $L$. But,
in the case of the Bose systems with broken phase symmetry, one has
to consider the GFs with odd numbers of operators as well. This
makes the quantum-field formalism for the superfluid Bose systems
more complicated in comparison with the analogous one for the Fermi
systems.

The two-point (single-particle) GFs are determined by the formulae
\begin{equation} \label{EQ78}
\begin{array}{ll}
\displaystyle{%
  G^{\alpha\alpha'}(x,\tau;x',\tau')=-\big\langle T_\tau\,\hat{\Phi}_\alpha(x,\tau)\,\hat{\Phi}_{\alpha'}(x',\tau') \big\rangle\,,  %
}\vspace{3mm}\\ %
\displaystyle{ %
  \tilde{G}^{\alpha\alpha'}(i\tau,i'\tau')=-\big\langle T_\tau\,\hat{\gamma}_{\alpha i}(\tau)\,\hat{\gamma}_{\alpha' i'}(\tau') \big\rangle\,. %
}
\end{array}
\end{equation}
These functions are $2\times 2$ matrices the ``isotopic'' space. The
components of GFs (\ref{EQ78}), which are diagonal in the isotopic
indices, are anomalous and different from zero only in the
superfluid state. On the contrary, the non-diagonal components
differ from zero in both the superfluid and normal states. To build
the perturbation theory, it is necessary to introduce the operators
in the Matsubara representation of interaction
\begin{equation} \label{EQ79}
\begin{array}{ll}
\displaystyle{%
  A_\alpha(\tau)= e^{\tau H_0}A_\alpha\,e^{-\tau H_0}\,.  %
}%
\end{array}
\end{equation}
Using these operators, we determine the temperature GFs in the
framework of the SCF model as
\begin{equation} \label{EQ80}
\begin{array}{ll}
\displaystyle{%
  G^{(0)\alpha\alpha'}\!(x,\tau;x'\tau')=-\big\langle T_\tau\,\Phi_\alpha(x,\tau)\,\Phi_{\alpha'}(x',\tau') \big\rangle_0\,,  %
}\vspace{3mm}\\ %
\displaystyle{ %
  \tilde{G}^{(0)\alpha\alpha'}\!(i\tau,i'\tau')=-\big\langle T_\tau\,\gamma_{\alpha i}(\tau)\,\gamma_{\alpha' i'}(\tau') \big\rangle_0\,. %
}
\end{array}
\end{equation}
Here, the averaging is carried out over the self-consistent state
with the statistical operator (\ref{EQ21}). The functions
(\ref{EQ78}) and (\ref{EQ80}) depend only on the difference of
``times'' $\tau-\tau'$.

To construct the perturbation theory, it is necessary to pass in
(\ref{EQ76}) from the averaging over the proximate state to the
averaging over the self-consistent state and to the operators in the
interaction representation. Thus, we get
\begin{equation} \label{EQ81}
\begin{array}{ll}
\displaystyle{%
  G(1,2,\ldots L)= i^L \frac{\big\langle T_\tau A(1)A(2)\ldots A(L)\,\sigma(\beta)\big\rangle_0}{\langle\sigma(\beta)\rangle_0 } \,,  %
}%
\end{array}
\end{equation}
where the temperature scattering matrix is
\begin{equation} \label{EQ82}
\begin{array}{ll}
\displaystyle{%
  \sigma(\beta)=T_\tau\exp\!{\bigg[ -g\int_0^\beta\!\! d\tau H_C(\tau)\bigg]}\,. %
}
\end{array}
\end{equation}
According to the connectivity theorem \cite{AGD,BKY} which also
remains valid in the given approach, the numerator in (\ref{EQ81})
can be represented in the form
\begin{equation} \nonumber
\begin{array}{ll}
\displaystyle{%
  \big\langle T_\tau A(1)A(2)\ldots A(L)\,\sigma(\beta)\big\rangle_0=  %
}\vspace{2mm}\\ %
\displaystyle{\hspace{3mm} %
  =\langle\sigma(\beta)\rangle_0 \big\langle T_\tau A(1)A(2)\ldots A(L)\,\sigma(\beta)\big\rangle_{0c}, %
}
\end{array}
\end{equation}
where the index ``c'' means the account of only connected diagrams.
As a result, the average of a temperature scattering matrix is
reduced in the nominator and denominator of (\ref{EQ81}) so that we
should account for only the connected diagrams in order to calculate
a GF. We note that the total thermodynamic potential of the system
is expressed in terms of the average of the temperature scattering
matrix over the self-consistent state. This average value can be
written in the form \cite{AGD,BKY}
\begin{equation} \nonumber
\begin{array}{ll}
\displaystyle{%
   \langle \sigma(\beta)\rangle_0 = \exp{ \!\left[ \sum_{n=0}^{\infty} \langle \sigma_n(\beta)\rangle_{0c} \right] }\,,   %
}
\end{array}
\end{equation}
whereas the total thermodynamic potential reads
\begin{equation}
\label{EQ83}
\begin{array}{ll}
\displaystyle{%
   \Omega=\Omega_0 - T \sum_{n=1}^{\infty} \langle \sigma_n(\beta)\rangle_{0c}\,.   %
}
\end{array}
\end{equation}

The Green's function can be represented as a series
\begin{equation} \label{EQ84}
\begin{array}{ll}
\displaystyle{%
   G(1,2,\ldots L)=\sum_{n=0}^{\infty} G^{(n)}\!(1,2,\ldots L)\,.   %
}
\end{array}
\end{equation}
The $n$-th order contributions to both the thermodynamic potential
and the GF are determined by the expressions
\begin{equation} \label{EQ85}
\begin{array}{ll}
\displaystyle{%
   \langle\sigma_n(\beta)\rangle_{0c} = \frac{g^n(-1)^n}{n!}\int_0^\beta\!\! d\tau_1\ldots  %
}\vspace{2mm}\\ %
\displaystyle{ \hspace{3mm} %
   \ldots\int_0^\beta\!\! d\tau_n \big\langle T_\tau\,H_C(\tau_1)\ldots H_C(\tau_n)\big\rangle_{0c}\,,   %
}
\end{array}
\end{equation}
\begin{equation} \label{EQ86}
\begin{array}{ll}
\displaystyle{%
   G^{(n)}\!(1,2,\ldots L)= \frac{g^n(-1)^n\,i^L}{n!}\int_0^\beta\!\! d\tau_1'\ldots  %
}\vspace{2mm}\\ %
\displaystyle{ \hspace{3mm} %
   \ldots\int_0^\beta\!\! d\tau_n'\big\langle T_\tau A(1)A(2)\ldots A(L) H_C(\tau_1')\ldots H_C(\tau_n')\big\rangle_{0c}.    %
}
\end{array}
\end{equation}
We recall that the correlation Hamiltonian consists of two terms
(\ref{EQ65}) and perform the further transformation of the
above-given formulae. It should be taken into account that only
those averages, which contain the even number of operators, are
different from zero in expressions (\ref{EQ85}) and (\ref{EQ86}).
With regard for this, formula (\ref{EQ85}) reads                                \vspace{-2mm} %
\begin{equation} \label{EQ87}
\begin{array}{ll}
\displaystyle{%
   \langle\sigma_{2n}(\beta)\rangle_{0c} = \frac{g^{2n}}{(2n)!}%
   \int_0^\beta\!\!d\tau_1\ldots\int_0^\beta\!\!d\tau_{2n}\times%
   }\vspace{2mm}\\ %
\displaystyle{ \hspace{0mm} %
   \times\bigg[ \Big\langle T_\tau H_C^{(3)}\!(\tau_1)\ldots H_C^{(3)}\!(\tau_{2n})\Big\rangle_{0c}+%
}\vspace{2mm}\\ %
\displaystyle{ \hspace{2mm} %
   + \Big\langle T_\tau H_C^{(4)}\!(\tau_1)\ldots H_C^{(4)}\!(\tau_{2n})\Big\rangle_{0c}+%
}\vspace{2mm}\\
\displaystyle{ \hspace{2mm} %
   + \sum_{l=1}^{n-1}C_{2n}^{2n-2l} \bigg\langle T_\tau
   \prod_{i=1}^{2n-2l}H_C^{(3)}\!(\tau_i)\!\!
   \prod_{j=2n-2l+1}^{2n}H_C^{(4)}\!(\tau_j) \bigg\rangle_{0c}\,\bigg], %
}
\end{array}
\end{equation}
for the even-order terms of the perturbation theory (for $n>1$) and %
\begin{equation} \label{EQ88}
\begin{array}{ll}
\displaystyle{%
   \langle\sigma_{2n+1}(\beta)\rangle_{0c} = -\frac{g^{2n+1}}{(2n+1)!}%
   \int_0^\beta\!\!d\tau_1\ldots\int_0^\beta\!\!d\tau_{2n+1}\times%
   }\vspace{2mm}\\ %
\displaystyle{ \hspace{0mm} %
   \times\bigg[ \Big\langle T_\tau H_C^{(4)}\!(\tau_1)\ldots H_C^{(4)}\!(\tau_{2n+1})\Big\rangle_{0c}+%
}\vspace{2mm}\\ %
\displaystyle{ \hspace{2mm} %
   + \sum_{l=0}^{n-1}C_{2n+1}^{2n-2l} \bigg\langle T_\tau
   \prod_{i=1}^{2n-2l}H_C^{(3)}\!(\tau_i)\!\!
   \prod_{j=2n-2l+1}^{2n+1}H_C^{(4)}\!(\tau_j) \bigg\rangle_{0c}\,\bigg], %
}
\end{array}
\end{equation}
for the odd-order ones (for $n\ge 1$). Here, $C_n^m$ are the
binomial coefficients. It follows from the properties of a normal
product that $\langle\sigma_1(\beta)\rangle_{0}=0$. Then, the
contribution of the SCF approximation corrections to the
thermodynamic potential becomes nonzero only in the second order in
perturbation. This means that the summation in expression
(\ref{EQ83}) starts from $n=2$. We write the expression for this
correction term separately:
\begin{equation} \label{EQ89}
\begin{array}{ll}
\displaystyle{%
   \langle\sigma_{2}(\beta)\rangle_{0c} = \frac{g^2}{2!}%
   \int_0^\beta\!\!d\tau_1\int_0^\beta\!\!d\tau_2\times%
   }\vspace{2mm}\\ %
\displaystyle{ \hspace{3mm} %
   \times\bigg[ \Big\langle T_\tau H_C^{(3)}\!(\tau_1) H_C^{(3)}\!(\tau_{2})\Big\rangle_{0}+%
                \Big\langle T_\tau H_C^{(4)}\!(\tau_1) H_C^{(4)}\!(\tau_{2})\Big\rangle_{0} \bigg].%
}
\end{array}
\end{equation}
The first-order contribution to the $L$-point GF has the form %
\newpage $\vspace{-6mm}$
\begin{equation} \label{EQ90}
\begin{array}{ll}
\displaystyle{%
   G^{(1)}(1,2,\ldots L)= %
}\vspace{2mm}\\ %
\displaystyle{ \hspace{3mm} %
   = -g\,i^L\!\int_0^\beta\!\!d\tau'\bigg[ \Big\langle T_\tau A(1) A(2)\ldots A(L)H_C^{(3)}\!(\tau') \Big\rangle_{0c} +%
}\vspace{2mm}\\ %
\displaystyle{ \hspace{3mm} %
   + \Big\langle T_\tau A(1) A(2)\ldots A(L)H_C^{(4)}\!(\tau') \Big\rangle_{0c}\, \bigg], %
}
\end{array}
\end{equation}
whereas the contribution of higher orders ($n\ge2$) is expressed by
the formula
\begin{equation} \label{EQ91}
\begin{array}{ll}
\displaystyle{%
   G^{(n)}(1,2,\ldots L)= \frac{g^n(-1)^n}{n!} i^L\! %
   \int_0^\beta\!\!d\tau_1'\ldots\int_0^\beta\!\!d\tau_n'\times%
   }\vspace{2mm}\\ %
\displaystyle{ \hspace{0mm} %
   \times\bigg[
   \Big\langle T_\tau A(1) A(2)\ldots A(L) H_C^{(3)}\!(\tau_1')\ldots H_C^{(3)}\!(\tau_n')\Big\rangle_{0c}+%
}\vspace{2mm}\\ %
\displaystyle{ \hspace{2mm} %
  +\Big\langle T_\tau A(1) A(2)\ldots A(L) H_C^{(4)}\!(\tau_1')\ldots H_C^{(4)}\!(\tau_n')\Big\rangle_{0c}+%
}\vspace{2mm}\\
\displaystyle{ \hspace{2mm} %
   + \sum_{l=1}^{n-1}C_{n}^{n-l} \bigg\langle T_\tau A(1) A(2)\ldots A(L)\times
}\vspace{2mm}\\
\displaystyle{ \hspace{2mm} %
   \times \prod_{i=1}^{n-l}H_C^{(3)}\!(\tau_i')\!\!
          \prod_{j=n-l+1}^{n}H_C^{(4)}\!(\tau_j') \bigg\rangle_{0c}\,\bigg]. %
}
\end{array}
\end{equation}
For a many-particle Bose system with the pair interaction, it is
enough to consider one-, two-, three-, and four-point GFs.

{\bf 7.} The formulae of the previous section are valid for the
representations of GFs in terms of the out-of-condensate field
operators and quasiparticle ones. First, we formulate the diagram
technique for the field GFs. We introduce the graphic designations
\begin{equation} \nonumber
\begin{array}{ll}
\displaystyle{ \hspace{0mm} %
   \scalebox{0.7}[0.7]{\includegraphics{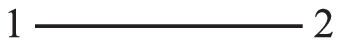}} \hspace{5mm}\textrm{for}\hspace{3mm}  G^{(0)}\!(1,2)=-\Phi^a(1)\Phi^a(2),  %
}\vspace{3mm}\\ %
\displaystyle{ \hspace{0mm} %
   \scalebox{0.7}[0.7]{\includegraphics{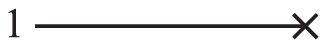}} \hspace{6mm}\textrm{for}\hspace{3mm}  i\chi(1),  %
}\vspace{3mm}\\ %
\displaystyle{ \hspace{0mm} %
   \scalebox{0.7}[0.7]{\includegraphics{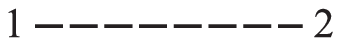}} \hspace{5mm}\textrm{for}\hspace{3mm}  \tilde{U}(1,2).  %
}
\end{array}
\end{equation}
The sign ``$\times$'' at the end of the line which corresponds to
the wave function of the Bose condensate means that no index
corresponds to this end. The construction of the diagram technique
is analogous to the case of Fermi particles
\cite{Poluektov3,Poluektov4}, with the single difference that is is
not necessary to show the direction of Green's lines in the
diagrams, and there exists an additional element -- the line of the
wave function of the Bose condensate.

\begin{figure*}[t!]
\centering %
\includegraphics[width = 1.6\columnwidth ]{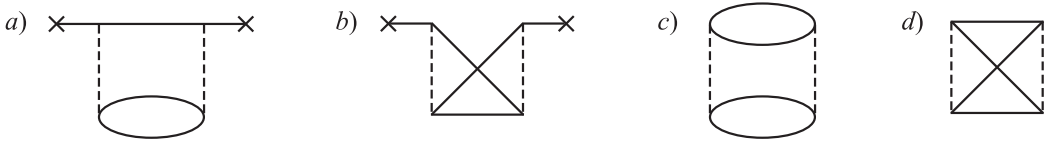} 
\caption{\label{F01}%
Second-order diagrams for the corrections to the temperature
scattering matrix in the field representation.
}%
\end{figure*}

We now calculate the second-order correction term to the temperature
scattering matrix which, according to (\ref{EQ83}), determines a
correction to the thermodynamic potential. Each of the first and
second terms in (\ref{EQ89}) corresponds to two nonequivalent
diagrams in Fig.\,\ref{F01}. The second-order contribution to the
temperature scattering matrix is determined by the formula %
\newpage $\vspace{-12mm}$

\begin{equation} \label{EQ92}
\begin{array}{ll}
\displaystyle{%
   \langle\sigma_{2}(\beta)\rangle_{0c} = \frac{g^2}{2!}\Bigg\{
   \frac{(-1)}{2^2}\int\!\!d1'd2'd1''d2''\times%
}\vspace{2mm}\\ %
\displaystyle{ \hspace{2mm} %
   \times\,\tilde{U}(1',2')\,\tilde{U}(1'',2'')\,\chi(1')\,\chi(1'')\,\times %
}\vspace{2mm}\\ %
\displaystyle{ \hspace{2mm} %
   \times\Big[ 2\,G^{(0)}\!(2',2'')\,G^{(0)}\!(\bar{2}',\bar{2}'')\,G^{(0)}\!(\bar{1}',\bar{1}'') + %
}\vspace{2mm}\\ %
\displaystyle{ \hspace{4mm} %
   + 4\,G^{(0)}\!(2',2'')\,G^{(0)}\!(\bar{2}',\bar{1}'')\,G^{(0)}\!(\bar{1}',\bar{2}'') \Big] +  %
}\vspace{2mm}\\ %
\displaystyle{ \hspace{2mm} %
   +\frac{1}{8^2}\int\!\!d1'd2'd1''d2''\,\tilde{U}(1',2')\,\tilde{U}(1'',2'')\,\times
}\vspace{2mm}\\ %
\displaystyle{ \hspace{2mm} %
   \times\Big[ 8\,G^{(0)}\!(1',1'')\,G^{(0)}\!(2',2'')\,G^{(0)}\!(\bar{2}',\bar{2}'')\,G^{(0)}\!(\bar{1}',\bar{1}'') + %
}\vspace{2mm}\\ %
\displaystyle{ \hspace{2mm} %
   + 16\,G^{(0)}\!(1',1'')\,G^{(0)}\!(2',2'')\,G^{(0)}\!(\bar{2}',\bar{1}'')\,G^{(0)}\!(\bar{1}',\bar{2}'') \Big]\Bigg\}. %
}
\end{array}
\end{equation}

\begin{figure}[t!]
\centering %
\includegraphics[width = 0.9\columnwidth ]{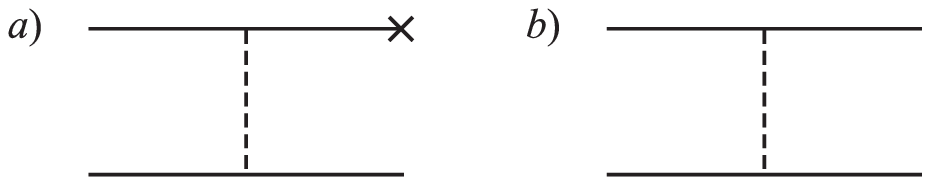} %
\caption{\label{F02}%
First-order diagrams for the corrections to the three-point ({\it
a}) and four-point ({\it b}) GFs in the field representation.
}\vspace{2mm}%
\end{figure}

We now turn to the consideration of the corrections to GFs. First,
we consider the first order of perturbation theory. In this order,
the corrections to the one- and two-point GFs are equal to zero. The
corrections to the three- and four-point GFs are described by the
diagrams $(a)$ and $(b)$, respectively, in Fig.\,\ref{F02}. In these
diagrams, the indices of the outer lines should be arranged by all
nonequivalent means; in our case, the number of such configurations
turns out to be three per each diagram. Analytically, the correction
term to the three-point GF has the form
\begin{equation} \label{EQ93}
\begin{array}{ll}
\displaystyle{%
   G^{(1)}\!(1,2,3)=-gi^3\!\int\!\!d1'd2'\,\tilde{U}(1',2')\,\chi(1')\,\times  %
   }\vspace{2mm}\\ %
\displaystyle{ \hspace{3mm} %
   \times\Big[ G^{(0)}\!(1,2')\,G^{(0)}\!(2,\bar{2}')\,G^{(0)}\!(3,\bar{1}') + %
}\vspace{2mm}\\ %
\displaystyle{ \hspace{5mm} %
              + G^{(0)}\!(1,2')\,G^{(0)}\!(2,\bar{1}')\,G^{(0)}\!(3,\bar{2}') + %
}\vspace{2mm}\\ %
\displaystyle{ \hspace{5mm} %
              + G^{(0)}\!(1,\bar{1}')\,G^{(0)}\!(2,2')\,G^{(0)}\!(3,\bar{2}') \Big], %
}
\end{array}
\end{equation}
whereas that to the four-point one reads 
\begin{equation} \label{EQ94}
\begin{array}{ll}
\displaystyle{%
   G^{(1)}\!(1,2,3,4)=-gi^4\!\int\!\!d1'd2'\,\tilde{U}(1',2')\,\times  %
   }\vspace{2mm}\\ %
\displaystyle{ \hspace{3mm} %
   \times\Big[ G^{(0)}\!(1,1')\,G^{(0)}\!(2,2')\,G^{(0)}\!(3,\bar{2}')\,G^{(0)}\!(4,\bar{1}') + %
}\vspace{2mm}\\ %
\displaystyle{ \hspace{5mm} %
              + G^{(0)}\!(1,1')\,G^{(0)}\!(2,2')\,G^{(0)}\!(4,\bar{2}')\,G^{(0)}\!(3,\bar{1}') + %
}
\vspace{2mm}\\ %
\displaystyle{ \hspace{5mm} %
              + G^{(0)}\!(1,1')\,G^{(0)}\!(2,\bar{1}')\,G^{(0)}\!(3,2')\,G^{(0)}\!(4,\bar{2}') \Big]. %
}
\end{array}
\end{equation}

\begin{figure}[h!]
\centering %
\includegraphics[width = 0.9\columnwidth ]{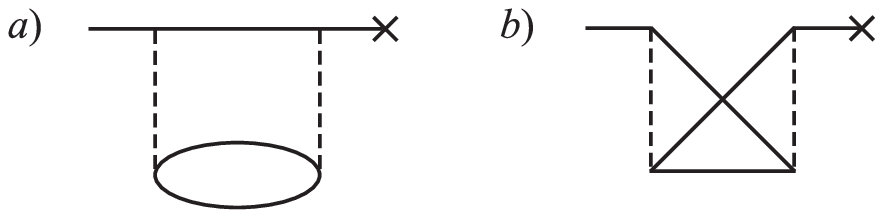} %
\vspace{-1mm}%
\caption{\label{F03}%
First-order diagrams for the corrections to the three-point ({\it
a}) and four-point ({\it b}) GFs in the field representation.
}%
\end{figure}
Consider the second order corrections. The corrections to the
one-point GF are described by the two diagrams in Fig.\,\ref{F03}
and have the forms
\begin{equation} \label{EQ95}
\begin{array}{ll}
\displaystyle{%
   G^{(2)}\!(1)= ig^2\!\int\!\!d1'd2'd1''d2''\,\tilde{U}(1',2')\,\tilde{U}(1'',2'')\,\chi(1')\,\times  %
   }\vspace{2mm}\\ %
\displaystyle{ \hspace{3mm} %
   \times\bigg[ \frac{1}{2}\,G^{(0)}\!(1,1'')\,G^{(0)}\!(2',2'')\,G^{(0)}\!(\bar{2}',\bar{2}'')\,G^{(0)}\!(\bar{1}',\bar{1}'') + %
}\vspace{2mm}\\ %
\displaystyle{ \hspace{8mm} %
              + G^{(0)}\!(1,1'')\,G^{(0)}\!(2',2'')\,G^{(0)}\!(\bar{2}',\bar{1}'')\,G^{(0)}\!(\bar{1}',\bar{2}'') \bigg]. %
}
\end{array}
\end{equation}
The second-order corrections to the higher order GFs can be
constructed in a similar manner.

The analysis of the formulae obtained shows that the diagram, which
describes the $n$-order contribution to the $L$-point GF, contains:

\noindent %
(a) $2n$ vertices connected in pairs by dashed lines (interaction
lines), each of which corresponds to the multiplier
$\tilde{U}\!(1',2')$;

\noindent %
(b) the internal solid lines (of the Green type), which correspond
to the multiplier $G^{(0)}\!(1',1'')$ and connect the vertices of
different dashed lines. We note that the Green's line cannot connect
the vertices of the same dashed line. In particular, its beginning
and end cannot belong to the same vertex;

\noindent %
(c) $L$ outer Green's lines, for which only one end is connected
with the interaction line vertex;

\noindent %
(d) the Bose condensate lines (with the sign ``$\times$'') which are
connected only with one vertex of a dashed line (the other end of
the interaction line cannot be connected with one more Bose
condensate line). The diagrams with the even number of outer
Greens's lines ($L=2S$) contain the even number of Bose condensate
lines. The diagrams with the odd number of outer Greens's lines
($L=2S+1$) contain the odd number of Bose condensate lines.

\begin{figure}[t!]
\centering %
\includegraphics[width = 0.9\columnwidth ]{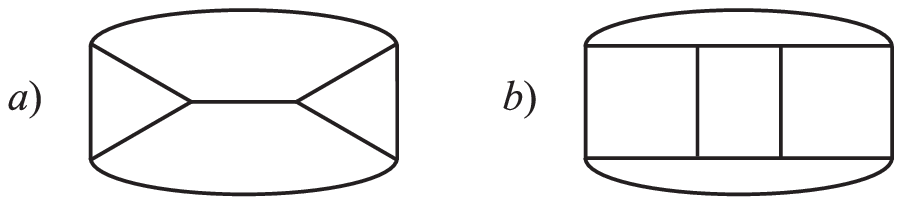} %
\vspace{0mm}%
\caption{\label{F04}%
Second-order diagrams for the corrections to the temperature
scattering matrix in the quasiparticle representation.
}%
\end{figure}

Thus, to calculate the $n$-order contribution to the $L$-point GF,
it is necessary:

\noindent %
(a) to depict all the topologically nonequivalent $n$-order
diagrams, i.e. those ones which don't turn into one another under
the permutations of the vertex indices of interaction lines;

\noindent %
(b) to associate the lines with their analytical expressions;

\noindent %
(c) to integrate over all the indices corresponding to the vertices
of interactions lines (the integration also includes the summation
over all the discrete indices);

\noindent %
(d) to perform such procedures: the index of the vertex of the
interaction line, which is included either in two GFs (when two
Green's lines converge into a vertex) or in the GF and the Bose
condensate function (when the GF and the Bose condensate line
converge into a vertex), has to be written once without the overbar,
and for the second time with the overbar;

\noindent %
(e) to put the multiplier $(-1)^n/2^k$ before the expression
obtained, where $n$ is the diagram order and $k$ is the number of
closed Green's lines in the diagram.

For the practical utilization of the diagram technique, the
frequency representation turns out to be more suitable. The Fourier
component of the $L$-point GF is handy to define as
\begin{equation} \label{EQ96}
\begin{array}{ll}
\displaystyle{%
   G(1,2,\ldots L; \omega_1,\omega_2,\ldots\omega_L)=  %
   }\vspace{2mm}\\ %
\displaystyle{ \hspace{0mm} %
   =\!\beta\Delta(\omega_1+\omega_2+\ldots +\omega_L)G(1,2,\ldots L; \omega_1,\omega_2,\ldots\omega_{L-1}), %
}
\end{array}
\end{equation}
where
\begin{equation} \label{EQ97}
\begin{array}{ll}
\displaystyle{%
   G(1,2,\ldots L; \omega_1,\omega_2,\ldots\omega_{L-1})=  %
}\vspace{2mm}\\ %
\displaystyle{ \hspace{3mm} %
   =\frac{1}{2^{L-1}}\int_{-\beta}^\beta\!d(\tau_1-\tau_2)\ldots\int_{-\beta}^\beta\!d(\tau_{L-1}-\tau_L)\times  %
}\vspace{2mm}\\ %
\displaystyle{ \hspace{3mm} %
   \times\,G(1,2,\ldots L; \tau_1-\tau_2,\ldots \tau_{L-1}-\tau_L)\times %
}\vspace{2mm}\\ %
\displaystyle{ \hspace{3mm} %
   \times\,e^{i\omega_1(\tau_1-\tau_2)+\ldots+i\omega_{L-1}(\tau_{L-1}-\tau_L) }. %
}
\end{array}
\end{equation}
In this case, in order to calculate the $L$-point GF, the rules of
the diagram technique have to undergo the following modifications:

\noindent %
(a) every Green's line is associated with the Fourier component
$G^{(0)}\!(1,2;\omega_n)$;

\noindent %
(b) every dashed line is associated with the potential
$\underline{\tilde{U}}(1,2)=U(x_1,x_2)(\delta_{\alpha_1\alpha_2}+\delta_{\alpha_1\bar{\alpha}_2})$;

\noindent %
(c) every dashed interaction line is associated with the multiplier
$\Delta(\omega_1'+\omega_2'+\omega_3'+\omega_4')$ or
$\Delta(\omega_1'+\omega_2'+\omega_3')$ (in the case where one of
the lines converging into a vertex is a Bose condensate line). Here,
$\omega_k'$ are the frequencies which correspond to the Green's
lines converging at the vertices of a given dashed line. In this
case, for all such GFs, the vertex indices for the interaction lines
have to be put either all at the first place or all at the second
place. If the order of indices is changed for a GF line, a sign is
to be changed in the above multipliers $\Delta$;

\noindent %
(d) the additional multiplier
$T^n\beta\Delta(\omega_1+\omega_2+\ldots+\omega_L)$, where $n$ is
the order of the diagram, emerges before the expression.

\begin{figure}[t!]
\centering %
\includegraphics[width = 0.7\columnwidth ]{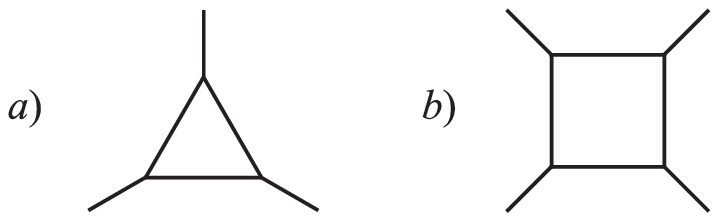} %
\vspace{-1mm}%
\caption{\label{F05}%
Second-order diagrams for the corrections to the temperature
scattering matrix in the quasiparticle representation.
}%
\end{figure}

Finally, we formulate the rules of the diagram technique in the
quasiparticle representation. We associate the matrix elements
(\ref{EQ74}) with a square and a triangle,
\vspace{0mm}%
\begin{equation} \nonumber
\hspace{0mm}
\begin{array}{ll}
\displaystyle{%
   \tilde{U}_{\bar{1}\bar{2}\bar{3}\bar{4}}\hspace{2mm}\textrm{---}\hspace{2mm}
   \scalebox{0.65}[0.65]{\includegraphics[bb = 28 490 85 510]{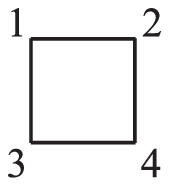}},\quad  %
   \tilde{U}_{\bar{1}\bar{2}\bar{3}}\hspace{2mm}\textrm{---}\hspace{2mm}
   \scalebox{0.65}[0.65]{\includegraphics[bb = 30 489 193 515]{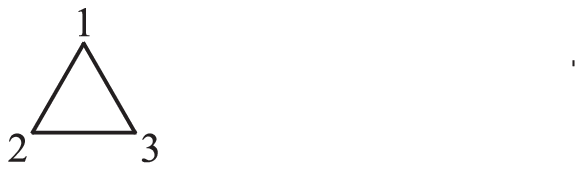}}\hspace{-26mm}, %
}\vspace{2.5mm} %
\end{array}
\end{equation}
and the solid Green's line with $G^{(0)}\!(i\tau,i'\tau')$. All the
indices of the square or the triangle correspond to the same time
parameter $\tau$. We recall that the wave function of the Bose
condensate is contained in the matrix element with three indices (in
the triangle). The $n$-order diagrams consist of $n$ squares and
triangles, whose vertices are connected by Green's lines by all
possible nonequivalent means. To calculate the $n$-order
contribution, we have to depict all topologically nonequivalent
diagrams, relate their elements to analytic expressions, and
integrate over the indices of the square and triangle vertices, as
well as the corresponding time parameters.

For the sake of illustration, we note that the second-order
contribution to the temperature scattering matrix is determined by
two diagrams in Fig.\,4, and the first-order contribution to the
three-point $(a)$ and four-point $(b)$ GFs -- by the diagrams shown
in Fig.\,5. In the quasiparticle representation, the diagrams are
simpler but the matrix elements of the interaction are much more
complex. As in the usual diagram technique, the block summation of
diagrams is allowable.

{\bf 8.} For the approach we developed, the self-energy and vertex
functions can be introduced, and the Dyson equations connecting
these functions can be written. The system of equations for the
one-point and two-point has the form
\begin{equation} \label{EQ98}
\begin{array}{ll}
\displaystyle{%
   \int\!\!d2\Big[\Omega(1,\bar{2})+g\,\Theta(1,\bar{2})\Big]G(2)\,+  %
   }\vspace{2mm}\\ %
\displaystyle{ \hspace{3mm} %
   +\,g\int\!\!d2d3\Big[ V^{(0)}(1,2,\bar{3})+\Lambda(1,2,\bar{3})\Big]G(3,2)= %
}\vspace{2mm}\\ %
\displaystyle{ \hspace{3mm} %
   = g\int\!\!d2d3\,V^{(0)}\!(1,2,\bar{3})\,G^{(0)}\!(3,2), %
}
\end{array}
\end{equation}
\begin{equation} \label{EQ99}
\begin{array}{ll}
\displaystyle{%
   \frac{\partial G(1,2)}{\partial\tau_1}+\sigma_{\bar{\alpha}_1}\int\!\!d3\Big[\Omega(1,\bar{3})+g\,\Theta(1,\bar{3})\,+  %
   }\vspace{2mm}\\ %
\displaystyle{ \hspace{3mm} %
   +\,g\Sigma(1,\bar{3})\Big]G(3,2) + g\sigma_{\bar{\alpha}_1} Z(1)\,G(2)=-\sigma_{\bar{\alpha}_1}\delta(1-\bar{2}),  %
}
\end{array}
\end{equation}
where $1=(x_1,\tau_1,\alpha_1)$ and $\bar{1}=(x_1,\tau_1,\bar{\alpha}_1)$. %
In Eqs.\,(\ref{EQ98}) and (\ref{EQ99}), the designations
\begin{equation} \label{EQ100}
\begin{array}{ll}
\displaystyle{%
   \Theta(1,2)=\tilde{U}(1,2)\Big[ G^{(0)}\!(1,2)-G(1,2)\Big] +   %
   }\vspace{2mm}\\ %
\displaystyle{ \hspace{3mm} %
   + \frac{1}{2}\,\delta(1-\bar{2})\int\!\!d3\,\tilde{U}(1,3)\Big[ G^{(0)}\!(3,\bar{3})-G(3,\bar{3})\Big],  %
}
\end{array}
\end{equation}
\begin{equation} \label{EQ101}
\begin{array}{ll}
\displaystyle{%
   \Lambda(1,2,3)= %
}\vspace{2mm}\\ %
\displaystyle{ \hspace{3mm} %
   = -\frac{1}{2}\,\tilde{U}(1,2)\int\!\!d1'd2'\,\Gamma(1',2',3)\,G(\bar{1}',1)\,G(\bar{2}',2),  %
}
\end{array}
\end{equation}
\begin{equation} \label{EQ102}
\begin{array}{ll}
\displaystyle{%
   Z(1)=\int\!\!d3d4\,V^{(0)}\!(1,3,\bar{4})\Big[ G(\bar{3},4)-G^{(0)}\!(\bar{3},4)\Big],  %
}
\end{array}
\end{equation}
\begin{equation} \label{EQ103}
\begin{array}{ll}
\displaystyle{%
   \Sigma(1,3)=\int\!\!d4\Big[ V^{(0)}\!(1,3,\bar{4})+V^{(0)}\!(1,4,\bar{3})\Big]G(4) +   %
}\vspace{2mm}\\ %
\displaystyle{ \hspace{0mm} %
   + \!\int\!\!d1'd2'd3'd4\,V^{(0)}\!(1,3',\bar{4})\,\Gamma(1',2',\bar{3})\,G(\bar{1}',3')\,G(\bar{2}',4)-  %
}\vspace{2mm}\\ %
\displaystyle{ \hspace{0mm} %
   -\frac{1}{2}\!\int\!\!d1'd2'd3'd4\,\tilde{U}(1,4)\,\Gamma(1',2',3',\bar{3})\times  %
}\vspace{2mm}\\ %
\displaystyle{ \hspace{45mm} %
   \times G(\bar{1}',1)\,G(\bar{2}',\bar{4})\,G(\bar{3}',4),  %
}
\end{array}
\end{equation}
\begin{equation} \label{EQ104}
\begin{array}{ll}
\displaystyle{%
   V^{(0)}\!(1,2,3)= %
}\vspace{2mm}\\ %
\displaystyle{ \hspace{3mm} %
   =-\frac{i}{2}\,\tilde{U}(1,2)\big[ 2\chi(2)\,\delta(1-3) + \chi(1)\,\delta(\bar{2}-3) \big] %
}
\end{array}
\end{equation}
are used. The three-point and four-point vertex functions read
\begin{equation} \nonumber
\begin{array}{ll}
\displaystyle{%
   G(1,2,3) = G(1,2)\,G(3) + G(1,3)\,G(2) + G(2,3)\,G(1)+ %
}\vspace{2mm}\\ %
\displaystyle{ \hspace{3mm} %
   + \int\!\!d1'd2'd3'\,\Gamma(1,2',3')\,G(\bar{1}',1)\,G(\bar{2}',2)\,G(\bar{3}',3), %
}
\end{array}
\end{equation}
\begin{equation} \label{EQ105}
\begin{array}{ll}
\displaystyle{%
   G(1,2,3,4) =  %
}\vspace{2mm}\\ %
\displaystyle{ \hspace{3mm}%
   = G(1,2)\,G(3,4) + G(1,3)\,G(2,4) + G(1,4)\,G(2,3)+ %
}\vspace{2mm}\\ %
\displaystyle{ \hspace{3mm} %
   + \int\!\!d1'd2'd3'd4'\,\Gamma(1,2',3',4')\times %
}\vspace{2mm}\\ %
\displaystyle{ \hspace{31mm} %
   \times\,G(\bar{1}',1)\,G(\bar{2}',2)\,G(\bar{3}',3)\,G(\bar{4}',4). %
}
\end{array}
\end{equation}

The one-point GF is associated with the self-energy functions by the
relation
\begin{equation} \label{EQ106}
\begin{array}{ll}
\displaystyle{%
   G(1,2) = G^{(0)}\!(1,2) + g\!\int\!\!d3d4\,G^{(0)}\!(1,\bar{3})\,\tilde{\Sigma}(3,\bar{4})\,G(4,2)+ %
}\vspace{2mm}\\ %
\displaystyle{ \hspace{3mm} %
   + g\!\int\!\!d3\,G^{(0)}\!(1,\bar{3})\,Z(3)\,G(2),  %
}
\end{array}
\end{equation}
where $\tilde{\Sigma}(1,2)=\Theta(1,2)+\Sigma(1,2).$ Expressions
(\ref{EQ101}) and (\ref{EQ103}) describe the relation of the
self-energy functions $\Lambda(1,2,3)$ and $\Sigma(1,3)$, on the one
hand, to the vertex functions $\Gamma(1,2,3)$ and $\Gamma(1,2,3,4)$,
on the other hand. These expressions are the analogs of the Dyson
equations for the many-particle Bose systems. As is seen, in the
Bose systems with the single-particle Bose condensate, the
additional vertex function $V^{(0)}\!(1,2,3)$, which originates from
both the interaction potential and the wave function of the Bose
condensate, emerges. We pay attention to the fact that the system of
equations (\ref{EQ98}) and (\ref{EQ99}), as well as the subsequent
relations, essentially differ from the Dyson equations obtained in
the Belyaev's approach \cite{Belyaev}. Contrary to the case
described in \cite{Belyaev}, this systems contains both the
two-point and one-point (condensate) GFs. The different structure of
the Dyson equations turns out to be very important. In particular,
it is on investigation of the Dyson equations that some general
statements about the character of the spectrum of excitations in the
Bose systems are based \cite{Bogolyubov3}.

It is known \cite{AGD} that the poles of a vertex function determine
the dispersion law for the collective excitations in a many-particle
system. This spectrum cannot be obtained as a result of the
calculation of the vertex function in any finite order of
perturbation theory. The four-point vertex function can be
represented as the sum of two functions, $\Gamma=\Gamma_1+\Gamma_2$.
One of these functions, $\Gamma_1$ is the sum of infinite
``stepwise'' series, whose terms are compact quadrilaterals
$\Gamma_k$ connected by the pairs of Green's lines which correspond
to exact GFs. The second function, $\Gamma_2$, contains all diagrams
which did not enter into $\Gamma_1$. Just the function $\Gamma_1$
gives rise to the appearance of the pole which corresponds to
collective excitations. It satisfies the relation
\begin{equation} \label{EQ107}
\begin{array}{ll}
\displaystyle{%
   \Gamma^{(1)}\!(1,2,3,4) = \Gamma_k(1,2,3,4)\,+ %
}\vspace{2mm}\\ %
\displaystyle{ \hspace{2mm} %
   +\!\int\!\!d5d6d7d8\,\Gamma_k(1,2,5,6)\,G(5,7)\,G(6,8)\,\Gamma^{(1)}\!(7,8,3,4).  %
}
\end{array}
\end{equation}

We note that the method of calculation of the dispersion law for
zero-sound collective excitations with the help of summation of the
infinite series of ``stepwise'' diagrams is well known in the theory
of Fermi systems \cite{AGD}. An analogous situation occurs also in
Bose systems. The assumption that the interparticle interaction is
weak makes it possible to substitute the exact GFs in (\ref{EQ107})
by their values in the zero approximation. This allows us to write
the equation which determines the dispersion law for the collective
excitations:
\begin{equation} \label{EQ108}
\begin{array}{ll}
\displaystyle{%
   1=\frac{U_0}{V}\,\sum_{{\bf p}}\frac{f(\varepsilon_{{\bf p}})-f(\varepsilon_{{\bf p}+{\bf k}})}
   {\omega-\varepsilon_{{\bf p}+{\bf k}}+\varepsilon_{{\bf p}}}.  %
}
\end{array}
\end{equation}
Here, $U_0$ is the interaction constant, $V$ is the volume,
$\varepsilon_{{\bf k}}$ is the dispersion law for a single-particle
excitation, and $f(\varepsilon_{{\bf k}})$ is the Bose distribution
function. Relation (\ref{EQ108}) yields the sound dispersion law for
the collective excitations, whose velocity
$c_0=\big(U_0n/m\big)^{1/2}$ ($m$ is the particle mass) doesn't
depend on temperature and is determined by both the particle number
density $n$ and interaction constant and is the same in the normal
and superfluid phases. It is the collective excitations that form
the linear part of the spectrum in many-particle Bose systems. The
independence of the linear part of the spectrum from temperature
(outside the hydrodynamic area) is confirmed in the experiments on
the inelastic scattering of slow neutrons in liquid $^4$He
\cite{BKKP,Kozlov}.

In this work, we have proposed a quantum-field method for the
theoretical description of many-particle Bose systems which are in
the states with broken symmetry. This approach is based on the
choice of the generalized model of self-consistent field as the
initial approximation. Such a choice of the basic approximation,
which is more realistic in comparison with the case of the model of
ideal Bose gas, makes it possible to avoid the difficulties emerging
in the available theory \cite{PS,N} and provides the opportunity to
investigate the spatially inhomogeneous states and, in particular,
the states with superfluid flows. In the basic approximation, the
spectrum of single-particle excitations of a Bose system is
calculated in the SCF model, whereas the spectrum of collective
excitations is determined by the poles of three- and four-point GFs
or vertex functions. The approach proposed does not contain any
assumptions, is based only on the general principles of quantum
mechanics and statistical physics, and is equally applicable for the
description of Fermi \cite{Poluektov3,Poluektov4} and Bose systems.

\pagebreak

\end{document}